\documentclass[preprint,nocomments,norevision]{ptephy_om}
\preprintnumber{RIKEN-iTHEMS-Report-26}
\revisionnum{3}

\usepackage{hyperref}
\usepackage{orcidlink}
\usepackage{enumitem}
\usepackage{graphicx}
\usepackage{tikz}
\usetikzlibrary{arrows.meta,calc,positioning}

\title{Two Languages for the Same Resonance:\\
From Nuclear Decay to Black-Hole Ringdown}

\author{Okuto Morikawa \orcidlink{0000-0002-0044-4491}}
\affil{Center for Interdisciplinary Theoretical and Mathematical Sciences (iTHEMS),
RIKEN, Wako 351-0198, Japan
\email{okuto.morikawa@riken.jp}
}

\begin{document}

\begin{abstract}
Quantum resonances and black-hole quasinormal modes (QNMs) are described by closely related mathematics: radiative boundary conditions, analytic continuation of Green functions or resolvents, poles on nonphysical sheets, and non-self-adjoint spectral representations. This article does not introduce that equivalence but reconstructs its genealogy. Nuclear resonance theory grew from radioactive decay, reaction cross sections, metastable compound states, and complex energies, whereas black-hole perturbation theory grew from spacetime stability and causal response.

We place the two developments on a common chronology, from Sommerfeld's radiation condition and early nuclear-decay theory through Regge--Wheeler perturbations, Vishveshwara's 1968 dissertation and 1970 papers, Press's 1971 quasi-normal terminology, and the independent Aguilar--Balslev--Combes theory of analytic dilation, then follow their later mathematical reunification in scattering-resonance theory.

Alongside the chronology, we give a working dictionary relating no-incoming and horizon-ingoing/infinity-outgoing conditions, pole energy and complex frequency, width and damping rate, residues and excitation factors, and nonresonant background with branch-cut and prompt contributions. The aim is to make the cross-disciplinary equivalence operational without flattening its history, and to distinguish mathematical equivalence from historical genealogy.

\end{abstract}

\maketitle
\tableofcontents

\section{Introduction: why a translation is needed}
\label{sec:intro}

A physicist trained in nuclear or atomic resonance theory and a physicist trained in black-hole perturbation theory can write essentially the same one-dimensional differential equation and then describe it in strikingly different language. The first may speak of a Gamow or Siegert state, a pole of the analytically continued $S$ matrix, a resonance width, and an outgoing-wave boundary condition. The second may speak of a quasinormal mode (QNM), a complex frequency, an ingoing condition at the event horizon, an outgoing condition at infinity, and a damped ringdown. A mathematical scattering theorist may instead speak of a pole of a meromorphically continued resolvent, while a practitioner of complex scaling may identify an angle-stable discrete eigenvalue of a non-self-adjoint deformation.

The proximity of these descriptions is no longer controversial. In suitable settings they are different realizations of the same scattering resonance. What is less often made explicit is why the terminology is different, which concepts appeared first, and which problem each community was trying to solve when a familiar mathematical structure entered its literature. The order matters. A historical account that starts from the modern equivalence and reads it backward risks turning Regge and Wheeler into resonance theorists, or Siegert into a precursor of black-hole physics. Neither description is historically useful. The more interesting story is that two research traditions approached the same analytic structure from different physical questions and only much later acquired a common language.

This article therefore has two aims. The first is chronological. We place the principal developments of quantum and nuclear resonance theory beside those of black-hole perturbation theory, concentrating on the period from the 1920s to the 1980s and then following the mathematical reunification in the 1990s and 2000s. The second is translational. At each stage we ask what was taken as primitive: a decaying state, a line shape, a causal boundary condition, a Green function, a pole, or an eigenvalue of a deformed operator.

Explicit cross-disciplinary identifications already exist in the literature. Kokkotas and Schmidt's 1999 review includes an appendix comparing the Schr\"odinger and wave equations for the same potential, notes that quantum mechanics encountered the same difficulties in defining resonances as black-hole mode theory, and retrospectively connects Gamow's alpha-decay problem with QNMs \cite{KokkotasSchmidt:1999}. De la Madrid later described Gamow states as the quasinormal modes of quantum systems \cite{DelaMadrid:2008GamowQNM}. Ashida, Gong, and Ueda place black-hole perturbations immediately after their discussion of quantum resonances and complex deformation, emphasizing the same radiative, non-square-integrable structure \cite{Ashida2020NonHermitian}. More recent black-hole work makes the vocabulary itself explicit: Motohashi calls the leaky QNM condition a Siegert boundary condition \cite{Motohashi:2024fwt}, while the black-hole spectroscopy review by Berti and collaborators identifies the nuclear analogue of a QNM wave function as a Gamow--Siegert state \cite{Berti:2025hly}.

These precedents make an important point clear: the identification of black-hole QNMs with quantum or scattering resonances is not new, and neither is the use of Gamow--Siegert language in black-hole physics. The contribution of the present article is therefore synthetic rather than a new QNM calculation or a new equivalence claim. Within the primary and secondary literature examined for this review, we have not identified a single article that simultaneously reconstructs the two genealogies on a common chronology while preserving their original problem framings, and then aligns boundary conditions, pole variables, widths, residues, and non-pole contributions as an operational dictionary for both communities. What appears to be missing is not the identification itself, but this historically ordered reconstruction together with the extended working dictionary. This bounded literature finding is not an assertion of absolute priority; the source and priority limitations stated below apply to it as well.

The central historical observation is a time lag followed by a remarkable near coincidence. The no-incoming-wave formulation of a resonance was explicit in nuclear reaction theory by the late 1930s, first as the energy-dependent Kapur--Peierls boundary problem and then as Siegert's self-consistent complex-energy pole condition \cite{KapurPeierls:1938,Siegert:1939}. Black-hole perturbation theory acquired its Schr\"odinger-like barrier equation in 1957 \cite{Regge:1957td}. In his 1968 doctoral dissertation, Vishveshwara explicitly imposed outgoing waves at infinity and purely ingoing waves at the Schwarzschild surface, related the resulting complex frequencies to resonance scattering and to poles of an $S(k)$ quantity introduced in analogy with a scattering-matrix element, and discussed their connection with causality \cite{Vishveshwara:1968Thesis}. His 1970 stability paper carried the same complex-frequency construction and its resonance/$S(k)$-pole interpretation into the journal literature, while his separate Nature paper exhibited the characteristic ringing seen in gravitational-wave scattering \cite{Vishveshwara:1970Stability,Vishveshwara:1970Scattering}. Press supplied the enduring quasi-normal terminology in 1971 \cite{Press:1971QNM}. In that same year, Aguilar--Combes and Balslev--Combes established the complex-dilation framework that gave a rigorous spectral realization of resonances in quantum mechanics \cite{Aguilar:1971ve,Balslev:1971vb}. These developments were not one research program. Their juxtaposition is nevertheless unusually revealing.

A second observation is methodological. The words ``ingoing at the horizon'' and ``outgoing at infinity'' can obscure the relation to Siegert's no-incoming condition. The horizon is the left asymptotic end of the tortoise-coordinate scattering problem. A wave that is ingoing with respect to the black-hole horizon is, from the point of view of the interaction region, a wave carrying flux out of the exterior system through its left channel. Thus a QNM is outgoing from the exterior through both asymptotic ends. The same equation can consequently be read in three dialects: causal black-hole response, quantum resonance scattering, and meromorphic spectral theory.

Our scope is deliberately narrower than a complete history of scattering theory, nuclear physics, or general relativity. We do not attempt to trace every precursor of the $S$ matrix, every formulation of unstable states, or every numerical method for QNMs.

A note on priority is necessary. This article reconstructs a conceptual genealogy from published papers and publicly accessible archival sources; it is not an exhaustive bibliometric study. Statements such as ``first,'' ``earliest,'' or ``already explicit'' therefore refer to the earliest documentary source identified in the present survey. We distinguish publication priority, archival evidence in dissertations, and the later introduction of standard terminology. Earlier lectures, correspondence, inaccessible manuscripts, or independently developed but undocumented formulations cannot be excluded. Where absolute priority cannot be established, we state only that a construction was explicit by a given date.

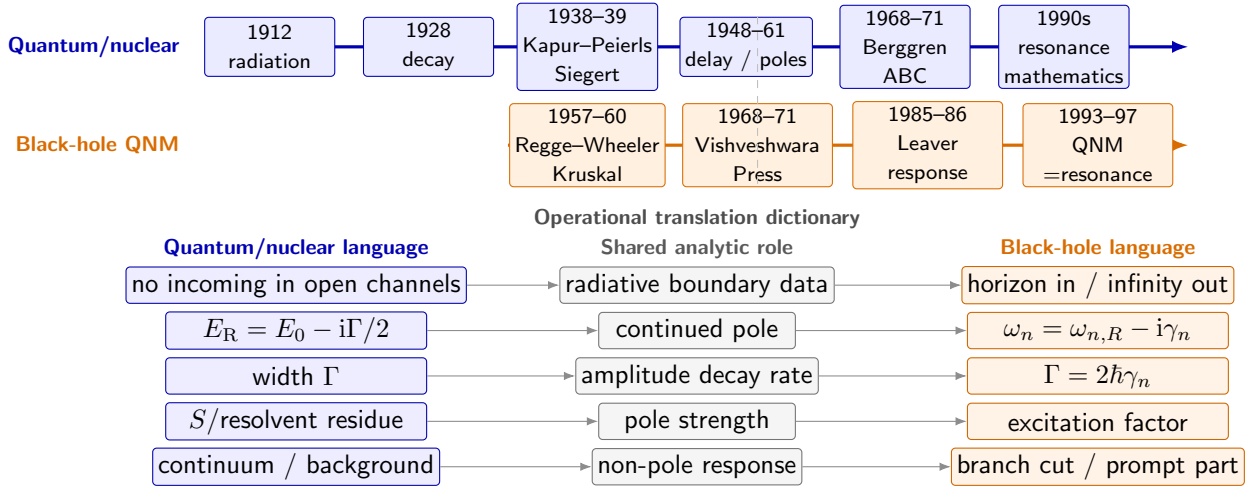
\begin{figure}[tb]
\centering
\resizebox{\textwidth}{!}{%
\begin{tikzpicture}[
  x=1cm,y=1cm,
  font=\sffamily\footnotesize,
  >=Latex,
  qnode/.style={draw=blue!70!black,fill=blue!8,rounded corners=1.5pt,
    font=\sffamily\scriptsize,align=center,minimum width=1.72cm,minimum height=0.78cm,inner sep=2pt},
  bnode/.style={draw=orange!85!black,fill=orange!12,rounded corners=1.5pt,
    font=\sffamily\scriptsize,align=center,minimum width=1.98cm,minimum height=0.78cm,inner sep=2pt},
  dleft/.style={draw=blue!65!black,fill=blue!7,rounded corners=1.5pt,
    align=center,minimum width=3.45cm,minimum height=0.48cm,inner sep=2pt},
  dmid/.style={draw=black!55,fill=black!4,rounded corners=1.5pt,
    align=center,minimum width=2.60cm,minimum height=0.48cm,inner sep=2pt},
  dright/.style={draw=orange!80!black,fill=orange!10,rounded corners=1.5pt,
    align=center,minimum width=3.45cm,minimum height=0.48cm,inner sep=2pt}
]
  \node[anchor=east,text=blue!70!black,font=\sffamily\bfseries\scriptsize]
    at (1.15,6.25) {Quantum/nuclear};
  \draw[blue!70!black,very thick,-{Latex[length=2.2mm]}] (1.35,6.25)--(14.35,6.25);
  \node[qnode] at (2.20,6.25) {1912\\radiation};
  \node[qnode] at (4.30,6.25) {1928\\decay};
  \node[qnode] at (6.40,6.25) {1938--39\\Kapur--Peierls\\ Siegert};
  \node[qnode] at (8.50,6.25) {1948--61\\delay / poles};
  \node[qnode] at (10.60,6.25) {1968--71\\Berggren\\ ABC};
  \node[qnode] at (12.70,6.25) {1990s\\resonance\\ mathematics};

  \node[anchor=east,text=orange!85!black,font=\sffamily\bfseries\scriptsize]
    at (1.15,4.95) {Black-hole QNM};
  \draw[orange!85!black,very thick,-{Latex[length=2.2mm]}] (5.35,4.95)--(14.35,4.95);
  \node[bnode] at (6.40,4.95) {1957--60\\Regge--Wheeler\\ Kruskal};
  \node[bnode] at (8.65,4.95) {1968--71\\Vishveshwara\\ Press};
  \node[bnode] at (10.90,4.95) {1985--86\\Leaver\\ response};
  \node[bnode] at (13.15,4.95) {1993--97\\QNM\\ $=$resonance};
  \draw[black!25,dashed] (8.65,4.44)--(8.65,6.76);

  \node[text=black!70,font=\sffamily\bfseries\scriptsize] at (7.85,3.98)
    {Operational translation dictionary};
  \node[text=blue!70!black,font=\sffamily\bfseries\scriptsize] at (2.55,3.58)
    {Quantum/nuclear language};
  \node[text=black!65,font=\sffamily\bfseries\scriptsize] at (7.85,3.58)
    {Shared analytic role};
  \node[text=orange!85!black,font=\sffamily\bfseries\scriptsize] at (13.15,3.58)
    {Black-hole language};

  \node[dleft]  (l1) at (2.55,3.10) {no incoming in open channels};
  \node[dmid]   (m1) at (7.85,3.10) {radiative boundary data};
  \node[dright] (r1) at (13.15,3.10) {horizon in / infinity out};
  \node[dleft]  (l2) at (2.55,2.50) {$\vsub{E}{R}=E_0-\rmi\Gamma/2$};
  \node[dmid]   (m2) at (7.85,2.50) {continued pole};
  \node[dright] (r2) at (13.15,2.50) {$\omega_n=\omega_{n,R}-\rmi\gamma_n$};
  \node[dleft]  (l3) at (2.55,1.90) {width $\Gamma$};
  \node[dmid]   (m3) at (7.85,1.90) {amplitude decay rate};
  \node[dright] (r3) at (13.15,1.90) {$\Gamma=2\hbar\gamma_n$};
  \node[dleft]  (l4) at (2.55,1.30) {$S$/resolvent residue};
  \node[dmid]   (m4) at (7.85,1.30) {pole strength};
  \node[dright] (r4) at (13.15,1.30) {excitation factor};
  \node[dleft]  (l5) at (2.55,0.70) {continuum / background};
  \node[dmid]   (m5) at (7.85,0.70) {non-pole response};
  \node[dright] (r5) at (13.15,0.70) {branch cut / prompt part};
  \foreach \n in {1,...,5}{
    \draw[black!45,-{Latex[length=1.5mm]}] (l\n.east)--(m\n.west);
    \draw[black!45,-{Latex[length=1.5mm]}] (m\n.east)--(r\n.west);
  }
\end{tikzpicture}%
}
\caption{Selected milestones in the two traditions and the translation dictionary used throughout this article. Horizontal spacing orders the landmarks but is not proportional to elapsed time. The arrows denote correspondence of analytic roles, not direct historical transmission}
\label{fig:timeline_dictionary}
\end{figure}

Figure~\ref{fig:timeline_dictionary} gives a compact synopsis of the article's distinctive combination: two selected historical sequences on one chronology and an operational translation dictionary. Its correspondences concern analytic roles, not evidence that one community transmitted its terminology directly to the other.

Within this scope, we follow the conceptual line needed to answer four questions:
\begin{enumerate}[label=(\roman*)\ ]
\item When did a radiation condition become part of the definition of a spectral problem?
\item When did complex energies and poles become the preferred language of unstable states?
\item How did black-hole stability calculations turn into a theory of damped resonances?
\item When and in what sense were black-hole QNMs identified with the resonances of mathematical scattering theory?
\end{enumerate}

The answer will also explain why complex scaling is a particularly effective translation device. It does not merely provide another numerical method. It converts the divergent outgoing-wave representation shared by Siegert states and QNMs into a square-integrable eigenvalue problem while retaining the pole and the rotated continuum. This is precisely the point at which the two languages become visibly the same spectral construction.

\section{One equation, three vocabularies}
\label{sec:dictionary_first}

Before following the chronology, it is useful to display the object whose history we are tracing. Consider the dimensionless one-dimensional stationary equation
\begin{equation}
\left[-\frac{\rmd^2}{\rmd x^2}+V(x)\right]\psi(x)=k^2\psi(x),
\label{eq:master}
\end{equation}
where $x\in\bbR$ is a scattering coordinate, $k$ is the asymptotic wave number, and the real potential $V(x)$ tends to zero at both ends. This common spatial equation must not be confused with a common dispersion relation. For a standard nonrelativistic Schr\"odinger problem,
\begin{equation}
E=\frac{\hbar^2 k^2}{2m},
\qquad
\Psi(t,x)=\rme^{-\rmi Et/\hbar}\psi(x),
\label{eq:schdisp}
\end{equation}
whereas for an asymptotically massless black-hole master field, in units with wave speed one, $k=\omega$ and we choose
\begin{equation}
\Psi(t,x)=\rme^{-\rmi\omega t}\psi(x).
\label{eq:timeconv}
\end{equation}
For real positive $k$, the elementary asymptotic waves are classified by their flux. A solution that carries flux away from the interaction region has the form
\begin{align}
\psi(x)&\sim \rme^{-\rmi kx},
& x&\longrightarrow-\infty,
\label{eq:leftout}\\
\psi(x)&\sim \rme^{+\rmi kx},
& x&\longrightarrow+\infty.
\label{eq:rightout}
\end{align}
The resonance condition is that a nonzero homogeneous solution satisfy both conditions simultaneously after analytic continuation in $k$ (or, equivalently, in the relevant spectral variable on a specified sheet).

In nuclear and quantum scattering, Eqs.~\eqref{eq:leftout} and \eqref{eq:rightout} are naturally called purely outgoing or no-incoming radiation conditions. In the historical literature they are associated, with qualifications discussed in Sec.~\ref{sec:siegert}, with the Kapur--Peierls and Siegert constructions \cite{KapurPeierls:1938,Siegert:1939}. If the energy pole is at
\begin{equation}
\vsub{E}{R}=E_0-\frac{\rmi}{2}\Gamma,
\qquad \Gamma>0,
\label{eq:quantumwidth}
\end{equation}
then the corresponding amplitude decays as $\rme^{-\Gamma t/(2\hbar)}$ and its probability as $\rme^{-\Gamma t/\hbar}$.

For a Schwarzschild black hole, $x$ is the tortoise coordinate $r_*$, with the event horizon at $r_*\to-\infty$ and spatial infinity at $r_*\to+\infty$. With the convention \eqref{eq:timeconv}, the QNM conditions are
\begin{align}
\psi(r_*)&\sim \rme^{-\rmi\omega r_*},
& r_*&\longrightarrow-\infty,
\label{eq:horingo}\\
\psi(r_*)&\sim \rme^{+\rmi\omega r_*},
& r_*&\longrightarrow+\infty.
\label{eq:infout}
\end{align}
The first is called \textit{ingoing at the future horizon}; the second is \textit{outgoing at infinity}. The sign difference is not a contradiction with the Siegert language. The null coordinates are
\begin{equation}
u=t-r_*,
\qquad
v=t+r_*.
\end{equation}
Then the horizon wave is $\rme^{-\rmi\omega v}$, regular and ingoing on the future horizon, whereas the infinity wave is $\rme^{-\rmi\omega u}$. From the exterior region, energy leaves through both endpoints. The event horizon is therefore a one-way outgoing channel of the exterior scattering system.

The same equivalence appears in the Green function. Let $f_-(x,k)$ satisfy the left outgoing condition and $f_+(x,k)$ the right outgoing condition. Their Wronskian,
\begin{equation}
W(k)
=f_-(x,k)\frac{\partial f_+(x,k)}{\partial x}
-\frac{\partial f_-(x,k)}{\partial x}f_+(x,k),
\end{equation}
is independent of $x$. The outgoing Green kernel is proportional to
\begin{equation}
G(x,x';k)
\propto
\frac{f_-(x_<,k)f_+(x_>,k)}{W(k)},
\label{eq:greenwronskian}
\end{equation}
where $x_<\equiv\min(x,x')$ and $x_>\equiv\max(x,x')$. A resonance wave number occurs when
\begin{equation}
W(k_n)=0.
\label{eq:Wzero}
\end{equation}
For an asymptotically massless black-hole problem $k_n=\omega_n$. Away from thresholds and branch points, and assuming analytically controlled asymptotics and no cancellation by the Green-kernel numerator, the following statements identify the same pole:
\begin{equation}
\begin{split}
&\text{no incoming wave in any open channel},\\
&\text{zero of an incoming Jost or connection coefficient},\\
&\text{zero of the outgoing Wronskian},\\
&\text{pole of the analytically continued Green function or resolvent}.
\end{split}
\label{eq:equiv}
\end{equation}
The location and sheet matter: zeros on other parts of the continued surface may instead describe bound, virtual, or antibound states. Complex scaling adds another representation of a resonance pole: it becomes an isolated eigenvalue of a non-self-adjoint deformed operator when the interaction is deformable and the pole is exposed from the rotated continuum \cite{Newton:1960,Aguilar:1971ve,Balslev:1971vb,Simon:73,Moiseyev:1998gjp,Zworski:2017}.

The rest of the article explains why these equivalences, transparent in modern notation, did not produce a common vocabulary from the beginning.

\section{Before resonance theory: radiation, decay, and line shapes, 1912--1936}
\label{sec:prehistory}

\subsection{Sommerfeld: outgoing radiation before quantum mechanics}

The oldest ingredient in our story is not a complex energy but an outgoing-wave condition. In 1912 Sommerfeld formulated the radiation condition in the study of the Green function for the Helmholtz equation \cite{Sommerfeld:1912radiation}. Its historical role was to select the physically radiating solution of an exterior boundary-value problem and thereby secure uniqueness. The later phrase ``Sommerfeld radiation condition'' should therefore be understood as part of classical mathematical physics, not as a device invented for quantum resonances. A useful historical reconstruction is given by Schot \cite{Schot:1992Sommerfeld}.

This point is conceptually important. A radiation condition first answers a question about \textit{which solution} of a wave equation represents a source radiating into an unbounded exterior. It need not define a discrete complex spectrum. The later resonance theories inherited this asymptotic selection rule and converted it into an eigenvalue condition by asking for a source-free solution that satisfies radiation conditions in all channels.

\subsection{Gamow and Gurney--Condon: radioactive decay as tunneling}

The quantum-mechanical part of the story begins from decay rather than scattering poles. In 1928 Gamow and, independently, Gurney and Condon explained alpha decay through barrier penetration \cite{Gamow:1928ZPhys,Gamow:1928Nature,GurneyCondon:1928Nature,GurneyCondon:1929}. The decisive physical content was that a state trapped for a long time inside a nuclear region can leak through a classically forbidden barrier. The observed decay law invited a description in terms of a lifetime and, eventually, a complex energy. What later generations call a Gamow state is therefore rooted in the physical problem of unstable nuclei, but the tunneling explanation itself has an independent Gurney--Condon line of discovery.

It is tempting to project the entire modern resonance formalism onto this starting point. That would hide the chronology. In 1928 the principal achievement was the tunneling explanation of nuclear decay, not the final operator-theoretic definition of a resonance as a pole of a meromorphically continued resolvent. The distinction between the historical problem and its later mathematical formulation will recur throughout this article.

\subsection{Breit--Wigner: the resonance as an experimentally visible line shape}

A second route to resonance came from reaction cross sections. Breit and Wigner's 1936 analysis of slow-neutron capture described resonance bands in terms of virtual nuclear excitation states and their damping widths \cite{BreitWigner:1936}. Here the emphasis shifts from the survival of an unstable state to the enhancement and width of a scattering probability.

The Breit--Wigner form became so successful that the word ``resonance'' is still often introduced phenomenologically as a peak with a width. Yet the later history repeatedly shows why the peak is not the most invariant object. Background interference can displace or even suppress a maximum; thresholds make widths energy dependent; and a broad pole need not generate an approximately Lorentzian profile. The line shape is an observable consequence of a pole together with a source, a background, and phase-space factors. The pole language arose partly because it separates those ingredients.

By the middle of the 1930s, therefore, two physical images were in place: a metastable state that decays and a reaction cross section that resonates. The next step was to define the unstable compound state directly by an asymptotic boundary condition.

\section{The no-incoming state: Kapur--Peierls and Siegert, 1938--1939}
\label{sec:siegert}

\subsection{Kapur and Peierls: the compound state as a boundary-value problem}

Kapur and Peierls' 1938 paper on the dispersion formula for nuclear reactions is a crucial bridge between phenomenology and the modern resonance boundary-value problem \cite{KapurPeierls:1938}. Their purpose was to formulate the compound-nucleus problem in coordinate space without relying on an overly restrictive perturbative picture. In the one-body prototype, the internal problem is matched at a finite channel radius to an external wave with the incident component removed. The boundary condition depends on the real scattering energy at which the collision problem is posed, so the resulting complex characteristic values and eigenfunctions are energy- and radius-dependent. They are not yet identical to the energy-independent pole states now usually called Siegert states \cite{Hategan:2020}.

This deserves emphasis because the terminology ``Siegert boundary condition'' can make the chronology appear cleaner than it was. The central physical operation---remove the incident wave and allow only decay into the external channel---was already explicit in Kapur and Peierls. Their construction also shows why a complex characteristic value can appear even though the underlying microscopic Hamiltonian is not being declared fundamentally non-Hermitian: non-self-adjointness enters through an open, energy-dependent boundary condition.

\subsection{Siegert: the eponym and the more portable formulation}

Siegert's short 1939 paper begins by referring directly to Kapur and Peierls and states that it follows a similar coordinate-space route while removing restrictive assumptions \cite{Siegert:1939}. Its decisive change is self-consistency: the exterior wave number belongs to the same complex energy being solved for. A state with only the outgoing component is then a singularity of the collision matrix, independent of an arbitrarily chosen real collision energy. The later resonance literature attached Siegert's name to this outgoing-wave pole eigenproblem, and the phrase ``Siegert state'' became standard in atomic, molecular, and nuclear physics \cite{Hategan:2020}.

The important historical lesson is not to choose one name and erase the other. A more accurate genealogy is
\begin{align}
\text{tunneling decay}
&\longrightarrow \text{Kapur--Peierls energy-dependent open problem},\\
&\longrightarrow \text{Siegert self-consistent pole state}.
\end{align}
Each step answered a different question. Gamow and Gurney--Condon supplied the physical mechanism of decay through a barrier. Kapur and Peierls made the open boundary condition part of a coordinate-space reaction problem. Siegert tied the outgoing wave to its own complex energy and thereby supplied the pole formulation that became an enduring label.

\subsection{The mathematical difficulty is already visible}

With the convention $\rme^{-\rmi Et/\hbar}$, a decaying resonance has $\im E<0$. Its outgoing momentum likewise lies below the real axis. The time dependence decays, but the stationary outgoing wave grows exponentially with distance. The resonance wave function is therefore not an ordinary $L^2$ eigenvector of the original self-adjoint Hamiltonian.

This tension is the technical seed of much of the subsequent literature. The resonance is physically meaningful, but its simplest stationary representative does not live in the Hilbert space in the same manner as a bound state. One can respond in several ways: define the resonance by an $S$-matrix or resolvent pole rather than by a vector; regularize the divergent state; enlarge the space of states; eliminate continuum channels to obtain an effective operator; or deform coordinates so that the outgoing wave becomes square integrable. The fact that all of these methods later converge on the same analytic pole is one of the great unifying achievements of resonance theory.

\section[What is a resonance? Spectra and effective Hamiltonians, 1948--1968]{What is a resonance? From continuous spectra to effective Hamiltonians, 1948--1968}
\label{sec:postwar}

The postwar development of resonance theory was not a single line. Several formulations attacked different aspects of the same problem: continuum coupling, time delay, complex eigenvalues, normalization, and completeness.

\subsection{Friedrichs and the instability of an embedded discrete state}

Friedrichs' 1948 analysis of perturbations of continuous spectra provided a controlled model in which a discrete level interacts with a continuum \cite{Friedrichs:1948}. In modern terminology, the Friedrichs model is a canonical laboratory for the conversion of an embedded or isolated discrete degree of freedom into an unstable resonance. Its importance lies less in reproducing one nuclear reaction than in making the discrete--continuum mechanism mathematically transparent.

This viewpoint later became central whenever a resonance was described as the remnant of a discrete state coupled to open channels. It also supplied a setting in which different resonance formalisms---effective Hamiltonians, analytic continuation, rigged Hilbert spaces, and complex scaling---could be compared explicitly.

\subsection[Eisenbud, Wigner, and Smith: resonance as delay]{Eisenbud, Wigner, and Smith: resonance as delay rather than merely a pole}

Eisenbud's 1948 Princeton dissertation related collision duration to the energy derivative of the scattering phase; Wigner's 1955 paper independently developed the causality and time-delay bounds while explicitly acknowledging Eisenbud's unpublished work \cite{Eisenbud:1948,Wigner:1955Delay}. With the common one-channel convention $S=\exp(2\rmi\delta)$, the phase delay is $2\hbar\,\rmd\delta/\rmd E$; factors depend on whether one differentiates the phase of $S$ or a partial-wave phase shift $\delta$. This real-frequency characterization remains important for black-hole physics because it detects resonant temporal response without first isolating a complex pole.

Smith generalized the idea in 1960 by introducing the lifetime matrix and showing its relation to the scattering matrix \cite{Smith:1960Lifetime}. In this language a metastable intermediate state is not only a complex pole. It is also detected by the excess time that a scattering process spends in the interaction region compared with free motion. Pole positions, line shapes, phase shifts, and delay times therefore became complementary ways of reading the same scattering process.

This is especially relevant to later black-hole applications. The QNM frequency by itself does not specify how strongly a given source excites the mode, and real-frequency quantities such as transmission phases and greybody factors can remain smooth even when a pole decomposition becomes ill-conditioned. The older scattering tradition had already developed a language in which response on the real axis and analytic poles are related but not identical.

\subsection[Feshbach: eliminated channels and effective Hamiltonians]{Feshbach: eliminated channels and an energy-dependent non-Hermitian operator}

Feshbach's unified theory of nuclear reactions gave another influential representation \cite{Feshbach1958UnifiedI,Feshbach1962UnifiedII}. Let $P$ and $Q$ be complementary orthogonal projectors, $P+Q=1$ and $PQ=0$, and write $H_{AB}\equiv AHB$ for $A$, $B\in\{P,Q\}$. Eliminating the $Q$ sector gives the energy-dependent effective operator
\begin{equation}
\vsub{H}{eff}(E)
=H_{PP}
+H_{PQ}\frac{1}{E-H_{QQ}+\rmi0}H_{QP}.
\label{eq:feshbach}
\end{equation}
Here $+\rmi0$ selects the outgoing boundary value on the physical cut. Its imaginary part represents flux lost from the selected sector into open $Q$ channels. The full theory can remain unitary and self-adjoint even though the reduced operator is energy dependent and non-Hermitian.

This distinction later became essential in the broader subject of non-Hermitian physics. A complex resonance eigenvalue need not imply fundamental nonunitarity. It can be a representation of a perfectly unitary scattering problem after boundary conditions, analytic continuation, or projection have been imposed.

\subsection[Peierls and Humblet--Rosenfeld: complex-energy resonances]{Peierls and Humblet--Rosenfeld: the resonance defined by complex-energy scattering}

Peierls' 1959 paper was explicit already in its title: ``Complex Eigenvalues in Scattering Theory'' \cite{Peierls:1959Complex}. The physical resonant states are solutions at complex energy for which one of the two asymptotic exponential components in every channel vanishes. The statement is remarkably close to the language one would now use to introduce a QNM as a non-self-adjoint boundary-value problem.

Humblet and Rosenfeld made the connection even sharper in 1961 \cite{HumbletRosenfeld:1961}. They defined resonant states as decaying states with complex energies characterized by natural boundary conditions expressing the absence of incoming waves in all channels. They then used the analytic behavior of the collision matrix and a Mittag--Leffler expansion to separate resonant and nonresonant contributions. By this stage the modern triad---complex resonance state, no-incoming boundary condition, meromorphic collision matrix---was already explicit in nuclear reaction theory.

A directly relevant historical synchronizer is Kruskal's 1960 maximal extension of the Schwarzschild metric \cite{Kruskal:1960}. By removing the coordinate singularity at the horizon, it supplied the regular geometric setting in which the causal horizon condition could later be stated; Vishveshwara's 1970 stability analysis explicitly used Kruskal coordinates. At that point quantum and nuclear resonance theory had already accumulated the concepts of complex energy, outgoing boundary conditions, scattering matrices, continuum coupling, time delay, and effective Hamiltonians, whereas black-hole QNMs had not yet emerged as a named subject. This is the scale of the historical offset between the two vocabularies.

\subsection[Zel'dovich and Berggren: normalization and completeness]{Zel'dovich and Berggren: how can a divergent state be normalized or completed?}

The outgoing Gamow--Siegert wave grows on the real coordinate axis. Zel'dovich's 1960 paper (published in English translation in 1961) introduced a Gaussian regularization for finite bilinear quantities, and Berggren's 1968 work developed resonant states in eigenfunction expansions \cite{Zel'dovich:1961,Berggren:1968zz}. The latter led to the Berggren completeness framework, in which selected resonant poles and a contour of continuum states appear together.

The conceptual point is more important here than the details of a particular normalization. The resonance cannot simply replace the continuum. A spectral representation that exposes resonant poles must retain an appropriate continuum contribution. This lesson reappears almost verbatim in black-hole response theory, where a QNM pole sum is accompanied by branch-cut or continuous contributions and by prompt/high-frequency pieces.

By the end of the 1960s, quantum resonance theory had therefore identified the central mathematical obstruction and several ways around it. Black-hole perturbation theory was about to rediscover the same open-wave structure from a completely different starting point.

\section{A black hole becomes a scattering problem, 1957--1970}
\label{sec:bhbeginnings}

\subsection{Regge--Wheeler: the question was stability, not resonance}

Regge and Wheeler's 1957 paper is a foundational event in black-hole perturbation theory \cite{Regge:1957td}. After decomposing nonspherical perturbations of the Schwarzschild geometry, they obtained the radial wave equation now associated with the Regge--Wheeler potential. In tortoise-coordinate form, it is a one-dimensional barrier-scattering equation of the same structural type as Eq.~\eqref{eq:master}.

From the viewpoint of modern scattering theory, it is difficult not to see a resonance problem. Historically, however, that is not what Regge and Wheeler set out to formulate. Their title states the concern: stability of the Schwarzschild solution. The central danger was an exponentially growing perturbation. The complex-frequency plane was therefore first interrogated for unstable modes rather than for damped poles governing a transient response.

This difference in question is one of the reasons the language diverged. Nuclear physicists asked how to describe an unstable intermediate state that decays into open channels. Relativists asked whether a spacetime background survives small perturbations and how waves behave near a horizon. Both questions lead to non-self-adjoint boundary conditions, but they assign different physical meaning to them.

\subsection{The tortoise coordinate and the causal meaning of the left channel}

For Schwarzschild spacetime, the tortoise coordinate maps the exterior region $2M<r<\infty$ to the full line,
\begin{equation}
-\infty<r_*<+\infty,
\end{equation}
with the event horizon at the left endpoint. The effective potential tends to zero at both ends. This converts the geometric perturbation problem into a scattering problem with two asymptotic channels.

The horizon channel is nevertheless not introduced as an abstract absorber. Regularity in coordinates crossing the future horizon fixes the causal branch. With the convention $\rme^{-\rmi\omega t}$, the regular wave behaves as $\rme^{-\rmi\omega(t+r_*)}$. Thus the statement ``ingoing at the horizon'' has a geometrical origin: it selects the solution regular on the future horizon and excludes radiation emerging from the past horizon.

This causal derivation is more than terminology. It explains why black-hole physicists did not need to borrow the phrase ``Siegert boundary condition'' in order to discover the same mathematics. The appropriate boundary condition followed directly from the global causal structure of the spacetime.

\subsection{Vishveshwara and Zerilli: from stability to radiative resonances}

Vishveshwara's 1968 doctoral dissertation on Schwarzschild stability contains a decisive formulation of the resonance problem \cite{Vishveshwara:1968Thesis}. After the stability analysis, he introduced complex frequencies under the boundary conditions of only outgoing waves at infinity and purely incoming waves at the Schwarzschild surface. He then related the purely outgoing condition to poles of a quantity $S(k)$ defined in analogy with a scattering-matrix element, interpreted the resulting enhancement as resonance scattering, and discussed the role of causality. In other words, the black-hole analogue of the outgoing resonance problem was already stated before the term ``quasi-normal mode'' became standard.

The associated analysis entered the journal literature in 1970 \cite{Vishveshwara:1970Stability}. The published paper retained the complex-frequency construction with outgoing radiation at infinity and ingoing radiation at the Schwarzschild surface, together with its resonance and scattering-matrix-pole interpretation. It also emphasized horizon regularity in Kruskal coordinates, ruled out the relevant exponentially growing perturbations, and noted the absence of a general Sturm--Liouville-type completeness theorem for the even-parity problem. Zerilli's 1970 work supplied the even-parity Schwarzschild master equation and its effective potential \cite{Zerilli:1970PRL,Zerilli:1970wzz}.

The complementary observational turn came in Vishveshwara's 1970 Nature paper on scattering gravitational radiation by a Schwarzschild black hole \cite{Vishveshwara:1970Scattering}. A wave packet scattered by the effective barrier produced characteristic damped oscillations determined by the black-hole geometry. In modern language, the time-domain response revealed the lowest QNMs.

A recent retrospective by Dadhich and Nayak situates these stability and scattering papers within Vishveshwara's broader black-hole program and their later reception \cite{DadhichNayak:2024}. It is used here as contextual secondary literature; the dates and technical priority statements above rest on the 1968 dissertation and the original 1970 papers.

A 2026 preprint by Besson and Jaramillo revisits Vishveshwara's scattering waveform in a hyperboloidal, non-self-adjoint formulation using a Keldysh QNM expansion \cite{BessonJaramillo:2026}. In their decomposition, the second peak correlates with the fundamental QNM and its first overtones, whereas the first even-parity peak is controlled by the algebraically special mode and a nearby branch cut. The result is a modern illustration of a central distinction in this review: identifying the poles is not the same as reconstructing the complete waveform, whose non-pole contributions must also be retained.

The sequence 1968--1970 therefore establishes both sides of the early black-hole story: an explicitly resonant complex-frequency boundary problem and its visible time-domain ringing. Press's later contribution was not the first statement of the pole condition, but the physical language that made these oscillations recognizable as the black hole's quasi-normal modes.

\section{1971: two independent turns}
\label{sec:1971}

The year 1971 is the most striking crossing point in the parallel chronology. On the black-hole side, Press interpreted the damped ringing as a family of quasi-normal oscillations \cite{Press:1971QNM}. On the quantum mathematical side, Aguilar--Combes and Balslev--Combes established the analytic-dilation framework that underlies the modern complex scaling method \cite{Aguilar:1971ve,Balslev:1971vb}. The two developments addressed different questions and belonged to different communities, but together they nearly complete the modern dictionary.

\subsection{Press: from a scattered waveform to quasi-normal frequencies}

Press studied long wave trains in time-domain evolutions and introduced the quasi-normal terminology that became standard in the subject \cite{Press:1971QNM,KokkotasSchmidt:1999}. His paper described a black hole as vibrating in a quasi-normal mode, but it did not present the full complex pole spectrum later associated with that phrase. The adjective is nevertheless revealing. A normal mode of a closed conservative system has a real frequency and a square-integrable or finite-energy mode function under the relevant boundary conditions. A black hole instead loses wave energy through the event horizon and to infinity. Its characteristic oscillations are therefore damped and are represented by complex frequencies.

In this language the black hole behaves like a bell, but not a perfectly isolated bell. The analogy emphasizes a characteristic spectrum of the object. It does not initially foreground the $S$-matrix pole or the analytic continuation of a self-adjoint exterior problem. That difference of emphasis helped QNM theory develop a vocabulary of modes, overtones, ringdown, and excitation amplitudes rather than one centered on Gamow states and nuclear dispersion formulae.

\subsection[Aguilar--Combes and Balslev--Combes: resonance as eigenvalue]{Aguilar--Combes and Balslev--Combes: turning a resonance into an eigenvalue}

The complex scaling problem begins from the spatial divergence noted in Sec.~\ref{sec:siegert}. For a short-range Schr\"odinger Hamiltonian
\begin{equation}
H=-\frac{\hbar^2}{2m}\nabla^2+V(x),
\end{equation}
consider the analytic dilation
\begin{equation}
x\longmapsto x\rme^{\rmi\theta}.
\end{equation}
Formally, the kinetic term transforms as
\begin{equation}
-\nabla^2\longmapsto -\rme^{-2\rmi\theta}\nabla^2.
\end{equation}
If an outgoing resonance momentum is
\begin{equation}
\vsub{k}{R}=|\vsub{k}{R}|\rme^{-\rmi\phi},
\qquad \phi>0,
\end{equation}
then along the rotated ray
\begin{equation}
\rme^{\rmi \vsub{k}{R}x\rme^{\rmi\theta}}
\end{equation}
has modulus $\exp[-|\vsub{k}{R}|x\sin(\theta-\phi)]$ and therefore decays for $\theta>\phi$, provided the deformation remains inside the analyticity sector of the interaction. The outgoing Gamow wave, divergent on the real axis, then becomes square integrable on the deformed contour.

Aguilar--Combes and Balslev--Combes made this picture into a theorem for suitable dilation-analytic interactions \cite{Aguilar:1971ve,Balslev:1971vb}. The continuous spectrum rotates by $2\theta$ into the lower half-plane in the energy variable $E\propto k^2$ (and, in a many-body problem, along rays issuing from the thresholds), while resonances uncovered between the physical axis and those rays appear as isolated eigenvalues independent of the dilation angle within their common analyticity domain. Simon soon extended and clarified the framework using quadratic forms and many-body resonance theory \cite{Simon1972BalslevCombes,Simon:73,Moiseyev:1998gjp}.

The central achievement is often summarized by saying that complex scaling ``rotates the continuum.'' That description is incomplete. The deeper statement is that matrix elements of the physical resolvent admit a meromorphic continuation and that the exposed resonance poles are represented by discrete eigenvalues of the deformed closed operator. The non-self-adjoint eigenproblem is therefore not a new definition disconnected from scattering. It is a representation of an already defined continued pole.

\subsection{Why the coincidence matters}

The juxtaposition is almost too neat. In one corner of physics, a damped black-hole response was being named as a quasi-normal spectrum. In another, the mathematical difficulty of representing outgoing quantum resonances was being solved by analytic dilation. Both concerned solutions that decay in time, grow in space on the original real coordinate, and are selected by radiation conditions. Yet there was no single shared program.

This coincidence is useful precisely because it prevents a triumphalist history. One should not say that the ABC theorem ``explained QNMs'' in 1971, nor that Press ``rediscovered Siegert states'' in any simple sociological sense. The accurate statement is more interesting: by 1971 two independent traditions had arrived at complementary halves of the modern resonance picture. One had a new physical system and a characteristic damped spectrum; the other had a rigorous spectral mechanism for outgoing resonances.

\section{From ringing to a radiative boundary-value problem, 1972--1986}
\label{sec:ringing_to_spectral}

\subsection{Goebel: temporary storage near the unstable photon orbit}

Goebel's 1972 note sharpened the physical interpretation of Press's ``vibrations'' \cite{Goebel:1972}. In the geometric-optics regime, the ringing can be understood in terms of gravitational waves temporarily stored near the unstable circular null orbit around the black hole. They are not material oscillations of a surface. Radiation leaks away both toward infinity and into the horizon.

In contemporary resonance language this is a barrier-top or trapped-set resonance. The potential barrier creates temporary confinement; the openness of the two channels gives a decay width. The photon-sphere picture and the nuclear metastable-state picture are not identical microscopically, but they organize the same dynamical structure: temporary localization plus leakage.

\subsection{Teukolsky: the rotating problem retains radiative asymptotics}

Teukolsky's separation of perturbation equations for rotating black holes extended QNM physics beyond the spherical Schwarzschild problem \cite{Teukolsky:1972my,Teukolsky:1973}. For a separated mode proportional to $\exp(-\rmi\omega t+\rmi m\varphi)$, where $m$ is the azimuthal number and $\vsub{\Omega}{H}$ the horizon angular velocity, the Kerr horizon wave number is shifted from $\omega$ to $\omega-m\vsub{\Omega}{H}$. Suppressing spin-dependent radial prefactors, the causal exponential is schematically
\begin{equation}
R(r)\sim
\rme^{-\rmi(\omega-m\vsub{\Omega}{H})r_*},
\qquad r_*\longrightarrow-\infty,
\end{equation}
while the asymptotically flat infinity condition remains outgoing. The superradiant regime makes it especially clear that the horizon is a physical scattering channel whose flux properties depend on the Killing frequency measured by the horizon generator.

This is already a warning against making the dictionary too literal. The abstract statement ``outgoing through all open channels'' survives, but the channel momentum need not be the same function of $\omega$ at each endpoint. In multi-channel or rotating problems, the correct translation is made at the level of flux and analytic boundary data, not by copying plane-wave exponents without context.

\subsection{Chandrasekhar--Detweiler: systematic QNM frequencies}

In 1975 Chandrasekhar and Detweiler made the named Schwarzschild QNM problem systematic and computed its characteristic frequencies \cite{ChandrasekharDetweiler:1975QNM}. Their abstract states the defining asymptotics as purely outgoing gravitational waves at infinity and purely ingoing waves at the horizon, and the problem is reduced to a one-dimensional short-range potential barrier. This was not the first occurrence of those asymptotics or their pole interpretation---both were already explicit in Vishveshwara's 1968 dissertation, with the associated stability results entering the journal literature in 1970---but it established the numerical complex-frequency spectrum as a central object in its own right.

At this point the mathematical equivalence is difficult to miss. If the Schwarzschild exterior is viewed as the system, the horizon and infinity are its two loss channels. ``Ingoing at the horizon'' means flux leaving the exterior to the left; ``outgoing at infinity'' means flux leaving to the right. The QNM is therefore a source-free solution with no incoming wave from either asymptotic channel.

Historically, however, the terminology remained gravitational. The QNM was a property of black-hole perturbations, and the central tasks were the computation of frequencies, the understanding of stability, and the relation to gravitational radiation. The fact that the boundary-value problem had a prehistory in nuclear reaction theory did not become the organizing narrative of the subject.

\subsection[Leaver: spectral decomposition of response]{Leaver: from complex frequencies to the spectral decomposition of response}

Leaver's 1985 continued-fraction method made accurate QNM computations possible for Schwarzschild and Kerr black holes \cite{Leaver:1985ax}. The following 1986 paper is even more important for the conceptual bridge \cite{Leaver:1986gd}. There Leaver constructed the radiative Green function and decomposed the time-domain response into a sum over QNM poles, an integral around a branch cut, and a high-frequency remnant of the free-space propagator.

In schematic notation,
\begin{equation}
\vsub{G}{ret}
=\vsub{G}{QNM}+\vsub{G}{cut}+\vsub{G}{direct}.
\label{eq:leaverdecomp}
\end{equation}
This is the black-hole counterpart of the resonance-expansion lesson already familiar from quantum scattering: pole terms do not by themselves constitute the entire response. The continuum or cut carries information that cannot be discarded, and a direct or prompt term may be needed as well.

The physical waveform acquires a corresponding temporal structure. A prompt response is followed, over an intermediate interval, by exponentially damped ringdown dominated by isolated QNM poles; at late times, low-frequency nonanalytic structure can produce the power-law behavior first analyzed systematically by Price \cite{Price:1972I,Ching:1995}. The decomposition is not merely a phenomenological sequence in time. It is the inverse-transform image of different pieces of the analytic structure of the Green function. Its precise validity depends on the source, observation point, function space, and contour deformation; it is not an unrestricted completeness theorem for QNM eigenfunctions \cite{Leaver:1986gd}.

By the mid-1980s, therefore, black-hole perturbation theory had independently assembled nearly all the ingredients that a scattering theorist would recognize: a radiative boundary-value problem, complex poles, residues, a Green function, and non-pole continuum contributions. What remained was an explicit mathematical identification of the terminology.

\section{Mathematical reunification: black-hole QNMs become resonances}
\label{sec:reunification}

\subsection{Bachelot and Motet-Bachelot: a rigorous scattering-theoretic synthesis}

A particularly important convergence occurred in the mathematical literature with Bachelot and Motet-Bachelot's 1993 paper, ``Les r\'esonances d'un trou noir de Schwarzschild'' \cite{BachelotMotetBachelot:1993}. Its significance goes beyond the title. The paper sets the Schwarzschild resonance problem inside a rigorous scattering-theoretic framework, relating time-dependent wave propagation to the corresponding stationary scattering problem and treating the complex frequencies through the analytic structure of the scattering operator. The black-hole problem is therefore not merely compared with resonance theory after the fact; its characteristic damped frequencies are incorporated into that theory as scattering resonances.

The paper also makes the terminological bridge explicit. The frequencies and mode functions called quasi-normal in the astrophysical literature are identified with the resonant objects of the scattering formulation. In this sense, 1993 is better viewed as a genuine synthesis than as a change of vocabulary: the causal wave problem, the stationary scattering problem, and the resonance interpretation are placed in one mathematical construction. This does not erase Vishveshwara's earlier resonance language or the preceding QNM literature; rather, it marks one of the earliest systematic embeddings of the black-hole problem into rigorous scattering-resonance theory.

\subsection[S\'a~Barreto and Zworski: geometric scattering and complex scaling]{S\'a~Barreto and Zworski: geometric scattering meets complex scaling}

S\'a~Barreto and Zworski's 1997 work advances this synthesis in a particularly revealing direction \cite{SaBarretoZworski:1997}. Their stated purpose is to apply methods of geometric scattering theory to the de~Sitter--Schwarzschild black hole, and they identify at the outset the ``resonances (or the quasi normal modes, in the terminology of Chandrasekhar).'' They prove a global meromorphic definition of the resonances in that setting and derive their high-angular-momentum pseudo-lattice structure. The emphasis has therefore shifted from establishing that black-hole complex frequencies can be regarded as resonances to studying their distribution with the machinery of geometric and semiclassical scattering theory.

For the present genealogy, an additional methodological link is especially significant. After reducing the radial problem to a one-dimensional semiclassical Schr\"odinger operator, S\'a~Barreto and Zworski obtain the meromorphic continuation directly by complex scaling and cite Aguilar--Combes as the original mathematical treatment \cite{Aguilar:1971ve,SaBarretoZworski:1997}. Thus one branch of the analytic-deformation theory developed for quantum resonances in 1971 enters explicitly into the black-hole resonance analysis. The coincidence emphasized in Sec.~7 consequently acquires a later documentary connection: Press's quasi-normal language and the Aguilar--Combes complex-scaling tradition arose independently in 1971, but by 1997 a black-hole QNM calculation was explicitly using the latter tradition to analyze the former objects.

The continuity from the 1993 synthesis is also visible in the documentary record. S\'a~Barreto and Zworski cite Bachelot and Motet-Bachelot's Schwarzschild-resonance work and acknowledge A. Bachelot for suggesting the de~Sitter model as especially suitable for the general methods of geometric scattering theory \cite{BachelotMotetBachelot:1993,SaBarretoZworski:1997}. The transition from 1993 to 1997 is therefore not merely a retrospective sequence imposed here: it traces an identifiable line from rigorous black-hole scattering theory to geometric resonance theory and complex scaling.

\subsection{Kokkotas and Schmidt: retrospective unification from the physics side}
\label{sec:kokkotas-retro}

An especially relevant bridge appeared in the 1999 Living Reviews article by Kokkotas and Schmidt \cite{KokkotasSchmidt:1999}. Its appendix, ``Schr\"odinger Equation Versus Wave Equation,'' compares the stationary Schr\"odinger problem and the separated wave equation for the same potential, explains resonances through poles of the analytically continued Green function, and remarks that quantum mechanics encountered the same conceptual difficulty in defining resonances as black-hole mode theory. It then retrospectively places Gamow's alpha-decay treatment at the beginning of the QNM story.

This is an important explicit identification, but it also illustrates the distinction emphasized in the present survey. As a modern mathematical classification, it is natural to place Gamow states and black-hole QNMs in the same family of resonant states. As a historical genealogy, however, that compression skips the intermediate changes of problem and definition: Gamow and Gurney--Condon addressed tunneling decay, Kapur and Peierls introduced an energy-dependent no-incident-wave boundary problem, Siegert imposed the self-consistent complex-energy pole condition, and the black-hole terminology arose independently from stability and radiative response. The 1999 statement is therefore best read as a retrospective unification, not as evidence that the later QNM concept or vocabulary was historically present in 1928.

\subsection[Vanzo and Zerbini: analytic dilation in QNM calculations]{Vanzo and Zerbini: analytic dilation becomes a QNM computational method}

Vanzo and Zerbini's 2004 analysis of multi-horizon black holes marks a further step in this convergence \cite{Vanzo:2004fy}. They formulate QNMs as scattering resonances associated with poles of a meromorphically continued resolvent and then use dilatation-analytic, or complex-scaling, methods to represent those resonances spectrally. The method is applied directly to the large-angular-momentum asymptotics of QNM frequencies for nonrotating multi-horizon geometries, with the Nariai limit serving as an exactly tractable check.

The historical significance is methodological rather than a claim of priority. Bachelot and Motet-Bachelot had already embedded the Schwarzschild problem in rigorous scattering theory, and S\'a~Barreto and Zworski had already used complex scaling within geometric black-hole resonance theory. Vanzo and Zerbini show the same analytic-deformation machinery operating explicitly inside a physics-facing QNM calculation. The two developments juxtaposed in 1971 are therefore connected not only retrospectively at the level of language: by 2004 the analytic-dilation tradition of quantum resonance theory had become an explicit computational route to black-hole QNM spectra.

\subsection{Dyatlov and the modern meromorphic definition}

The later microlocal literature made the operator-theoretic definition still more precise. For scalar waves on Kerr--de~Sitter spacetime, Dyatlov defined QNMs as poles of a meromorphic family of operators and showed that this rigorous definition agrees with the heuristic one used in the physics literature \cite{Dyatlov:2011QNM}. His global discreteness and exponential-decay results in that paper assume sufficiently slow rotation. The language of resonance expansions, trapped sets, and energy decay then places black-hole perturbations directly inside contemporary scattering theory.

From this viewpoint the conceptual hierarchy is clean. The physical retarded problem determines a resolvent or Green operator in a half-plane where it is unambiguously defined. That object is continued meromorphically in an appropriate functional setting. Its poles are the resonances. A QNM wave function is a residue state or generalized eigenfunction associated with a pole; its precise realization depends on the chosen representation.

\subsection{The translation becomes explicit in later reviews}

The cross-disciplinary identification did not remain confined to mathematical scattering theory. De la Madrid's 2008 nuclear-physics paper describes Gamow states directly as quasinormal modes of quantum systems and connects their resonance amplitude with the Breit--Wigner form \cite{DelaMadrid:2008GamowQNM}. Ashida, Gong, and Ueda's broad review of non-Hermitian physics moves directly from quantum resonances and complex deformation to black-hole perturbations, emphasizing the same radiative boundary conditions, failure of square integrability, complex eigenvalues, and finite lifetimes \cite{Ashida2020NonHermitian}. From the black-hole side, Motohashi explicitly uses the term ``Siegert boundary condition'' for QNMs \cite{Motohashi:2024fwt}, and the 2026 black-hole spectroscopy review makes the nuclear analogy equally explicit by identifying the corresponding resonance wave function as a Gamow--Siegert state \cite{Berti:2025hly}.

These examples establish that the equivalence itself is part of the modern literature. What they do not by themselves provide is the two-sided genealogy developed here: how the nuclear and black-hole constructions arose from different questions, which intermediate definitions were historically distinct, and how the translation extends beyond the boundary condition and pole position to widths, residues, excitation factors, and non-pole response. Modern reviews of black-hole QNMs consequently use a vocabulary much closer to scattering and non-Hermitian spectral theory than the early literature did \cite{KokkotasSchmidt:1999,Berti:2009kk,Berti:2025hly}. Yet disciplinary habits remain. ``Quasinormal mode'' is still the natural phrase when the emphasis is gravitational ringdown, whereas ``resonance'' is more common when the emphasis is resolvent continuation or geometric scattering.

\section{A working dictionary between the two traditions}
\label{sec:dictionary}

The history is most useful if it produces a practical dictionary. The translations below are not mere word substitutions; each carries assumptions about time conventions, asymptotic channels, and analytic sheets.

\subsection{Boundary conditions}

In quantum scattering one often writes
\begin{equation}
\text{Siegert condition: no incoming wave in any open channel}.
\end{equation}
For an asymptotically flat static black hole, the corresponding statement is
\begin{equation}
\text{future-horizon ingoing}
+\text{spatial-infinity outgoing}.
\end{equation}
These are the same flux condition when the exterior region is regarded as the open system. The apparent verbal mismatch occurs because ``ingoing'' at the horizon is defined relative to the horizon, while ``outgoing'' in Siegert language is defined relative to the interaction region.

The equivalence must be stated with the time convention. For $\rme^{-\rmi\omega t}$, a stable decaying QNM satisfies
\begin{equation}
\im\omega_n<0.
\end{equation}
Changing the time convention reverses all signs in the spatial exponents and the half-plane of decay.

\subsection{Complex energy, complex frequency, and width}

Quantum resonance theory often parameterizes a pole as
\begin{equation}
\vsub{E}{R}=E_0-\frac{\rmi}{2}\Gamma.
\end{equation}
Black-hole theory usually writes
\begin{equation}
\omega_n=\omega_{n,R}-\rmi\gamma_n,
\qquad \gamma_n>0.
\end{equation}
These parameters belong to their respective time-evolution generators. If one introduces the quantum angular frequency $\Omega\equiv E/\hbar$ and compares it with the black-hole frequency $\omega$, then
\begin{equation}
\vsub{\Omega}{R}=\frac{E_0}{\hbar}-\rmi\frac{\Gamma}{2\hbar}
\quad\Longrightarrow\quad
\Gamma=2\hbar\gamma_n
\label{eq:widthfactor2}
\end{equation}
under the formal identification $\Omega=\omega$. The factor of two is an elementary but persistent source of mistranslation: $\Gamma/\hbar$ is the probability-decay rate, whereas $\gamma_n=-\im\omega_n$ is an amplitude-decay rate.

This temporal dictionary is distinct from the stationary radial dictionary. In the Schr\"odinger equation $E=\hbar^2k^2/(2m)$, whereas for the massless black-hole equation used here $k=\omega$ and the radial spectral parameter is $\omega^2$. Thus one may identify the outgoing wave number $k$ between the two radial problems, or compare the time generators $E/\hbar$ and $\omega$, but one must not impose both identifications as a physical dispersion relation.

\subsection{\texorpdfstring{$S$}{S}-matrix pole, Green-function pole, and Wronskian zero}

A resonance can be identified as a pole of a continued scattering matrix, a pole of a continued resolvent or Green function, or a zero of an appropriate Jost function or Wronskian. In one channel these formulations are usually related by elementary identities such as Eq.~\eqref{eq:greenwronskian}. In black-hole calculations the Wronskian of a horizon-ingoing solution and an infinity-outgoing solution often plays the role of the Jost denominator.

The important caveat is that analytic continuation is part of the definition. A self-adjoint Hamiltonian on the physical Hilbert space has real spectrum. A resonance is not inserted into that spectrum as an additional complex $L^2$ eigenvalue. It appears after continuation through the continuous-spectrum cut, or as an eigenvalue of a deformed representation that realizes the same continuation.

\subsection{Residue, excitation factor, and observable amplitude}

A pole location does not determine a measured ringdown amplitude. Near a simple pole,
\begin{equation}
G(\omega)
=\frac{A_n}{\omega-\omega_n}+\vsub{G}{reg}(\omega),
\label{eq:residue}
\end{equation}
where $A_n$ is an operator-valued residue. For a source $|S\rangle$ and an observation functional $\langle O|$, the measured pole amplitude is
\begin{equation}
a_n=\langle O|A_n|S\rangle.
\end{equation}
In black-hole language, excitation factors and source-dependent excitation coefficients play roles analogous to resonance residues and channel couplings in scattering theory. A normalized QNM waveform is representation dependent; the source-to-observer residue is the invariant response datum.

\subsection{Ringdown, background, and continuum}

The quantum phrase ``resonance plus nonresonant background'' translates naturally into the black-hole decomposition of prompt response, QNM ringdown, and tail. Leaver's Eq.~\eqref{eq:leaverdecomp} is the cleanest historical bridge. The QNM sum is a pole contribution, the late-time tail is tied to nonanalytic continuum or branch-cut structure, and the prompt term contains high-frequency/direct propagation not represented by a finite set of poles.

This is why the question ``Are QNMs complete?'' is potentially misleading unless the function space, time interval, source class, observation region, and contour decomposition are specified. A pole expansion can be exact in one representation or domain and incomplete in another; Leaver's branch-cut and high-frequency terms are a concrete counterexample to an unrestricted pure-pole expansion \cite{Leaver:1986gd}. The correct object of comparison is the full causal response.

\subsection{Photon-sphere QNMs and barrier-top resonances}

In the eikonal regime, for the usual decoupled fields and under the assumptions of the geometric-optics correspondence, the connection between QNMs and unstable null geodesics makes the translation especially concrete \cite{Goebel:1972,Cardoso:2009}. The photon sphere produces temporary trapping near the top of an effective potential barrier; leakage to the horizon and infinity supplies the imaginary part of the frequency. In quantum language these are barrier-top resonances. The correspondence is an asymptotic statement and is not universal for every coupled field or modified-gravity perturbation, but Goebel's 1972 interpretation remains an early physical explanation of why a black-hole QNM behaves like a metastable scattering state.

\section{Why the vocabularies diverged}
\label{sec:why_diverge}

If the mathematical structure is so close, why did the two communities not use the same words from the beginning? The chronology suggests several reasons.

\subsection{They began from different observables}

Nuclear resonance theory was driven by decay rates, reaction cross sections, compound-nucleus levels, and lifetimes. Its characteristic quantities were energies and widths. Black-hole perturbation theory was driven by spacetime stability and gravitational radiation. Its characteristic quantities were mode frequencies, damping times, waveforms, and horizon fluxes.

The different observables encouraged different metaphors. ``Metastable state'' is natural when an excited nucleus exists temporarily before decaying. ``Ringdown mode'' is natural when a perturbed spacetime emits a damped waveform. The same pole can support both descriptions, but neither vocabulary is neutral about what is physically salient.

\subsection{The boundary condition had a different justification}

In the nuclear problem, the outgoing condition was introduced to define a decaying compound state or a pole of a scattering amplitude. In black-hole physics, the horizon condition is dictated by causal regularity on the future event horizon. No physical signal emerges from the future horizon into the exterior in a retarded problem. The infinity condition likewise expresses the absence of radiation sent in from infinity when one studies a free ringdown.

Thus the black-hole researcher can derive the QNM boundary conditions without ever asking how unstable states were defined in nuclear physics. The mathematics converges because both problems are open-wave systems, not because one necessarily borrowed the other's formalism.

\subsection{The self-adjoint starting point was more visible in quantum mechanics}

Quantum resonance theory developed against the background of the spectral theorem for self-adjoint Hamiltonians. The appearance of a complex ``eigenvalue'' therefore created an immediate conceptual problem: where does the state live, what is its norm, and how can it coexist with real spectrum? This tension directly motivated regularization, rigged Hilbert spaces, effective Hamiltonians, and complex scaling.

Black-hole perturbation theory often began instead from a hyperbolic initial-value problem and its separated radial equation. The use of complex frequencies did not create the same pedagogical clash with a previously declared self-adjoint Hamiltonian. The main question was whether the frequency-domain boundary problem correctly represented the causal evolution. As a result, the Hilbert-space status of a QNM could remain secondary to its role in the Green function and waveform.

\subsection{Disciplinary pathways and evidentiary limits}

A disciplinary explanation is plausible, but it should not be overstated. The primary sources assembled here establish distinct problem framings and local vocabularies; they do not by themselves quantify the degree of social or institutional separation among nuclear physicists, relativists, astrophysicists, and mathematical scattering theorists. A stronger claim about conference networks, textbook readerships, or direct channels of transmission would require a dedicated citation-network, prosopographical, or archival study beyond the scope of this survey.

What the documentary record does show is enough for the more limited point needed here. Nuclear-reaction papers organized the subject around compound states, collision matrices, widths, and open channels \cite{KapurPeierls:1938,Siegert:1939,Feshbach1958UnifiedI,HumbletRosenfeld:1961}; the early black-hole papers organized it around stability, horizon regularity, scattered waveforms, and quasi-normal oscillations \cite{Regge:1957td,Vishveshwara:1968Thesis,Vishveshwara:1970Stability,Vishveshwara:1970Scattering,Press:1971QNM}. Later mathematical work could then recognize both as instances of scattering resonance theory \cite{BachelotMotetBachelot:1993,SaBarretoZworski:1997}. A mathematically transferable construction therefore need not carry its original name across disciplinary boundaries.

This is precisely why historical genealogy and mathematical equivalence must be separated. It is legitimate to say that the Chandrasekhar--Detweiler QNM boundary problem is mathematically of Siegert type. It is not legitimate to infer from that equivalence that the black-hole authors were historically working inside the Siegert tradition unless the documentary record shows such a connection.

\section{Complex scaling as a translation device}
\label{sec:complex_scaling}

Complex scaling is especially useful for communication between the two subjects because it makes their common difficulty visible in one equation. On the outgoing resonance sheet, the relevant asymptotic wave number has $\im k<0$, so a QNM or Siegert wave grows exponentially at large $|x|$ on the real axis. A complex deformation rotates the asymptotic coordinate into a sector where the same analytic solution decays.

For a simple uniform scaling $x=y\rme^{\rmi\theta}$, Eq.~\eqref{eq:master} becomes schematically
\begin{equation}
H_\theta\psi_\theta
=
\left[
-\rme^{-2\rmi\theta}\frac{\rmd^2}{\rmd y^2}
+V(y\rme^{\rmi\theta})
\right]\psi_\theta
=
k^2\psi_\theta.
\label{eq:cs}
\end{equation}
The original outgoing boundary condition has been converted into an $L^2$ condition on a non-self-adjoint operator. The continuum rotates while an exposed resonance remains stationary with respect to $\theta$ inside the allowed analytic wedge \cite{Aguilar:1971ve,Balslev:1971vb,Simon:73}.

The transformation therefore supplies a literal dictionary:
\begin{equation}
\begin{split}
\text{Siegert/QNM outgoing solution}
&\longleftrightarrow
\text{square-integrable eigenvector of }H_\theta,\\
\text{continued resolvent pole}
&\longleftrightarrow
\text{angle-stable discrete eigenvalue},\\
\text{physical continuum}
&\longleftrightarrow
\text{rotated continuum of }H_\theta.
\end{split}
\end{equation}
This correspondence is powerful because it does not discard the continuum. In a finite basis, the rotated continuous spectrum appears as a sequence of deformation-dependent pseudostates, whereas a genuine exposed resonance is comparatively stable under changes of scaling angle and basis. Both pieces contribute to the reconstructed response.

Complex-scaling and geometric-resonance methods entered black-hole analysis well before current finite-basis implementations. As discussed in Sec.~\ref{sec:reunification}, S\'a~Barreto and Zworski used complex scaling in their 1997 geometric resonance analysis, and Vanzo and Zerbini subsequently made analytic dilation an explicit tool in a black-hole QNM calculation \cite{SaBarretoZworski:1997,Vanzo:2004fy}. This prehistory matters for priority: later implementations extend an established black-hole resonance program at the computational level rather than introducing complex scaling into the subject for the first time.

One should not overstate the result. The classical ABC theorem was proved for specified classes of dilation-analytic Schr\"odinger operators. A black-hole master equation introduces additional analytic structure: the tortoise map is logarithmic, horizons and singularities constrain admissible contours, Kerr equations can be frequency dependent, and asymptotically flat problems possess low-frequency branch structure responsible for tails. Applying complex scaling to a black-hole problem therefore requires a problem-specific analytic deformation, a declared spectral variable (for Eq.~\eqref{eq:master}, $\lambda=k^2$), and control of the relevant cuts. It does not by itself constitute a universal black-hole ABC theorem.

That limitation is scientifically productive. It identifies the precise place where a translation between disciplines becomes a research problem rather than a change of notation. The invariant target is the same continued response pole; the admissible representation depends on the analytic geometry of the operator.

Complex scaling is not the only representation that regularizes QNM eigenfunctions. In the hyperboloidal formulation, the future horizon and future null infinity are incorporated geometrically into the foliation, and QNMs become regular eigenfunctions of a non-self-adjoint time-evolution generator. This provides a complementary geometric realization of the same resonant data rather than an analytic coordinate deformation~\cite{Ansorg:2016ztf,PanossoMacedo:2024nkw}.

\section{What is not identical}
\label{sec:notidentical}

A good dictionary should state its failures as clearly as its successes. Several common black-hole problems lie outside the simplest two-ended, short-range Siegert model.

\subsection{Asymptotically anti-de~Sitter boundary conditions}

For asymptotically anti-de~Sitter spacetimes, the outer boundary is not ordinary radiative infinity. Reflecting, Dirichlet, Neumann, or mixed boundary conditions can be physically relevant depending on the field and holographic problem. Such QNMs remain poles of a retarded response, but the phrase ``purely outgoing at infinity'' is no longer the correct translation \cite{Berti:2009kk}.

\subsection{Massive fields and quasibound states}

A massive field can support exponentially decaying spatial behavior at infinity rather than an outgoing oscillatory wave. The resulting quasibound states mix the language of bound states and resonances. The channel structure and sheet choice must be specified before applying a Siegert dictionary \cite{Berti:2009kk}.

\subsection{Kerr superradiance}

In rotating spacetimes the horizon wave number is $\omega-m\vsub{\Omega}{H}$, and the horizon flux can change sign in the superradiant regime. ``Outgoing from the exterior'' remains a useful causal phrase, but ordinary intuition based on equal asymptotic momenta is insufficient. The correct analytic boundary condition is tied to the horizon generator and to the chosen physical sheet \cite{Teukolsky:1973,Berti:2009kk}.

\subsection{Long-range tails and branch points}

The Schwarzschild tortoise coordinate differs logarithmically from the areal radius at infinity, and the inverse-power/logarithmic asymptotics are not those of a compactly supported or exponentially decaying one-dimensional barrier. Low-frequency branch structure produces the Price tail and complicates both resolvent continuation and global complex deformation \cite{Price:1972I,Ching:1995,Leaver:1986gd}. The universal resonance idea survives, but the asymptotic comparison dynamics and deformation must be adapted to the actual long-range structure; an unqualified identification with the Coulomb problem would be misleading.

\subsection{A QNM is not synonymous with every non-Hermitian eigenvalue}

Finally, neither community benefits from identifying ``resonance'' with an arbitrary complex eigenvalue of a finite non-Hermitian matrix. A physical resonance is anchored to a parent scattering or response problem. The matrix may be an effective Hamiltonian, a complex-scaled discretization, a continued-fraction representation, a hyperboloidal generator, or a Galerkin approximation; its complex spectrum is meaningful only insofar as it represents the appropriate analytic data of the parent problem. Non-normality also makes eigenvalues sensitive to perturbations, so convergence and pseudospectral stability are distinct questions.

There is a useful historical coda to this warning. Nollert's 1996 step-potential study found QNM frequencies to be extremely sensitive to small changes of the underlying potential, while Nollert and Price later examined how completeness and excitation behave in QNM systems \cite{Nollert:1996rf,NollertPrice:1999}. Jaramillo, Panosso Macedo, and Al Sheikh recast this older sensitivity problem in 2021 using the pseudospectrum of a non-self-adjoint hyperboloidal formulation \cite{Jaramillo:2020tuu}. They found the slowest-decaying Schwarzschild QNM stable under perturbations that respect the asymptotic structure, while sufficiently high-frequency perturbations can strongly displace the overtones. This development should not be read as a pathology created by complex scaling: the 2021 construction is a different non-self-adjoint representation. Its relevance to the present genealogy is more general. Once a resonance is represented spectrally by a non-self-adjoint operator, one must distinguish the representation-independent scattering or response pole of a fixed parent problem from the conditioning and perturbation sensitivity of a particular operator realization.

\section{Chronology: the two languages on one time axis}
\label{sec:chronology}

The parallel history can now be compressed into a single chronology. The purpose of this list is not to assign exclusive priority for every concept, but to show the changing questions and terminology.

\begin{description}[leftmargin=18mm,style=nextline]
\item[1912] Sommerfeld formulates the radiation condition for exterior wave problems, selecting outgoing solutions at infinity \cite{Sommerfeld:1912radiation,Schot:1992Sommerfeld}.

\item[1928] Gamow and, independently, Gurney and Condon explain nuclear alpha decay through quantum tunneling, establishing the physical prototype of a metastable quantum state \cite{Gamow:1928ZPhys,Gamow:1928Nature,GurneyCondon:1928Nature,GurneyCondon:1929}.

\item[1936] Breit and Wigner connect nuclear resonances with observable reaction line shapes and widths \cite{BreitWigner:1936}.

\item[1938] Kapur and Peierls define compound-nucleus states through energy-dependent coordinate-space boundary conditions with no incident wave, producing complex characteristic values \cite{KapurPeierls:1938}.

\item[1939] Siegert makes the outgoing wave number self-consistent with the complex eigenenergy and identifies the states with collision-matrix singularities; the outgoing pole boundary condition later becomes associated with his name \cite{Siegert:1939,Hategan:2020}.

\item[1948] Friedrichs gives a controlled spectral model of a discrete state coupled to a continuum \cite{Friedrichs:1948}; Eisenbud's dissertation formulates collision time delay through the energy derivative of scattering phase \cite{Eisenbud:1948}.

\item[1955] Wigner relates the energy derivative of scattering phase to causality and delay, acknowledging Eisenbud's earlier unpublished result \cite{Wigner:1955Delay}.

\item[1957] Regge and Wheeler reduce Schwarzschild perturbations to a one-dimensional barrier equation. Their question is spacetime stability, not the construction of a resonance theory \cite{Regge:1957td}.

\item[1958] Feshbach develops an effective-Hamiltonian theory in which eliminated open channels generate an energy-dependent optical operator \cite{Feshbach1958UnifiedI}.

\item[1959] Peierls explicitly formulates physical resonance states as complex-energy scattering solutions with one asymptotic exponential removed in each channel \cite{Peierls:1959Complex}.

\item[1960] Smith introduces the lifetime matrix, connecting metastable lifetimes directly to the scattering matrix \cite{Smith:1960Lifetime}.

\item[1960--1961] Kruskal gives the maximal extension of the Schwarzschild metric in coordinates regular at the horizon \cite{Kruskal:1960}. On the resonance side, Zel'dovich introduces Gaussian regularization of unstable states in a 1960 Russian paper whose English translation appears in 1961 \cite{Zel'dovich:1961}, while Humblet and Rosenfeld characterize resonant states by no-incoming conditions and meromorphic collision-matrix expansions \cite{HumbletRosenfeld:1961}. Black-hole QNMs are not yet a named subject.

\item[1968] Berggren develops resonance expansions that keep selected resonant states together with a continuum contribution \cite{Berggren:1968zz}. In the same year, Vishveshwara's doctoral dissertation explicitly formulates Schwarzschild complex frequencies with outgoing radiation at infinity and purely ingoing radiation at the horizon, and relates them to resonance scattering and to poles of an $S(k)$-type scattering quantity introduced in analogy with a scattering-matrix element \cite{Vishveshwara:1968Thesis}.

\item[1970] Zerilli supplies the even-parity Schwarzschild master equation \cite{Zerilli:1970PRL,Zerilli:1970wzz}. Vishveshwara's stability paper carries the same complex-frequency, resonance-scattering, and $S(k)$-pole construction into the journal literature, while his separate Nature paper observes characteristic damped ringing in gravitational-wave scattering \cite{Vishveshwara:1970Stability,Vishveshwara:1970Scattering}.

\item[1971] Press introduces the quasi-normal terminology for black-hole ringing \cite{Press:1971QNM}. Independently, Aguilar--Combes and Balslev--Combes establish the analytic-dilation theory that represents resonances as discrete eigenvalues of a deformed operator \cite{Aguilar:1971ve,Balslev:1971vb}.

\item[1972--1973] Goebel interprets black-hole ringing as temporary storage near the unstable null orbit \cite{Goebel:1972}; Teukolsky separates perturbations of rotating black holes \cite{Teukolsky:1972my}; Simon develops the mathematical foundations of dilation-analytic resonance theory \cite{Simon1972BalslevCombes,Simon:73}.

\item[1975] Chandrasekhar and Detweiler systematize the named Schwarzschild QNM boundary problem and compute its characteristic complex frequencies \cite{ChandrasekharDetweiler:1975QNM}.

\item[1985--1986] Leaver develops an analytic/continued-fraction QNM representation and then decomposes the Schwarzschild causal response into QNM poles, a branch-cut integral, and a high-frequency remnant \cite{Leaver:1985ax,Leaver:1986gd}.

\item[1993] Bachelot and Motet-Bachelot explicitly study the ``resonances of a Schwarzschild black hole'' in mathematical scattering language \cite{BachelotMotetBachelot:1993}.

\item[1996] Nollert exposes strong sensitivity of QNM frequencies to small changes of a model potential, anticipating the later question of pseudospectral stability \cite{Nollert:1996rf}.

\item[1997] S\'a~Barreto and Zworski state the identification in direct terminology: the resonances are the quasi-normal modes of the physics literature, using geometric resonance and complex-scaling methods \cite{SaBarretoZworski:1997}.

\item[1999] Kokkotas and Schmidt explicitly compare the Schr\"odinger resonance problem with black-hole QNMs and retrospectively place Gamow's alpha-decay treatment at the beginning of the QNM story \cite{KokkotasSchmidt:1999}. In the same year, Nollert and Price investigate completeness and excitation in QNM systems, sharpening the distinction between the existence of complex frequencies and their role in reconstructing physical response \cite{NollertPrice:1999}.

\item[2004] Vanzo and Zerbini formulate black-hole QNMs as scattering resonances and meromorphically continued resolvent poles, and use dilatation-analytic methods directly to derive asymptotic QNM spectra for multi-horizon black holes \cite{Vanzo:2004fy}. This makes analytic deformation an explicit computational tool within black-hole QNM physics.

\item[2008] De la Madrid describes Gamow states as the quasinormal modes of quantum systems and relates their resonance amplitude to the Breit--Wigner form \cite{DelaMadrid:2008GamowQNM}.

\item[2011] Dyatlov gives a rigorous meromorphic-operator definition of scalar Kerr--de~Sitter QNMs and proves its agreement with the heuristic physics definition; the global discreteness and decay results stated there assume slow rotation \cite{Dyatlov:2011QNM}.

\item[2020--2021] Ashida, Gong, and Ueda review quantum resonances, complex deformation, and black-hole perturbations within a common non-Hermitian framework \cite{Ashida2020NonHermitian}. Jaramillo, Panosso Macedo, and Al Sheikh introduce pseudospectral analysis into black-hole QNM stability, distinguishing the robustness of the fundamental mode from the strong perturbation sensitivity of overtones in their Schwarzschild setting \cite{Jaramillo:2020tuu}.

\item[2025--2026] Motohashi explicitly calls the black-hole QNM radiation condition a Siegert boundary condition \cite{Motohashi:2024fwt}; the black-hole spectroscopy review by Berti and collaborators identifies the nuclear analogue of a QNM wave function as a Gamow--Siegert state \cite{Berti:2025hly}.

\end{description}

Two intervals summarize the historical geometry. The first is 1938--1939, when the no-incoming complex-energy resonance problem was established in nuclear reaction theory. The second is 1968--1971: Vishveshwara's dissertation formulated the black-hole resonance boundary problem, the 1970 papers brought the stability analysis and characteristic ringing into print, and Press supplied the quasi-normal name in 1971. The interval between the two formulations is roughly three decades. Yet 1971 simultaneously marks the mathematical maturation of quantum complex scaling. The histories are offset and synchronized at the same time.

\section{How to read the older papers without anachronism}
\label{sec:historiography}

The technical equivalence creates a special historiographical danger. Modern notation is powerful enough to erase the historical sequence. Three distinctions help avoid that mistake.

\subsection{A later unification is not an earlier motivation}

Regge and Wheeler can be rewritten as studying a Schr\"odinger operator with a barrier, but their paper should not be described simply as an early resonance calculation. Their stated problem was stability. Likewise, Gamow's alpha-decay argument can be embedded in modern pole theory, but its historical force was the explanation of nuclear tunneling. A modern equivalence tells us how results fit together now; it does not tell us what problem an author believed they were solving then.

The retrospective tracing of QNMs back to Gamow in Kokkotas and Schmidt's 1999 review, discussed in Sec.~\ref{sec:kokkotas-retro}, is a useful example: mathematically illuminating, but historiographically compressed.

\subsection{Names can postdate the underlying operation}

The phrase ``Siegert boundary condition'' is useful, but Kapur and Peierls had already used a no-incident-wave, energy-dependent coordinate-space construction; Siegert's self-consistent complex-energy pole condition was an additional step. Conversely, ``quasi-normal mode'' was introduced after Vishveshwara had described the radiative complex-frequency pole problem in his dissertation and had displayed characteristic ringing in the 1970 scattering paper. The history of a name and the history of an operation are not identical.

\subsection[Mathematical identity and physical interpretation]{Mathematical identity can coexist with different physical interpretation}

A pole of a continued Green function can be interpreted as a decaying compound-nucleus state, a temporarily trapped barrier wave, or a black-hole ringdown frequency. The pole is the same type of analytic object; the source, observable, asymptotic geometry, and physical interpretation differ. Good translation preserves both levels rather than reducing one field to the vocabulary of the other.

This is also why the phrase ``black holes are just quantum resonances'' would be unhelpful. Black-hole perturbation theory contains geometric structures with no direct analogue in the elementary nuclear barrier problem: horizons, gauge constraints, superradiance, spin-weighted fields, and spacetime asymptotics. The useful claim is narrower and stronger: \textit{the linear QNM spectral problem belongs to the same general theory of scattering resonances once its causal channels and analytic continuation are specified}.

\section{The modern synthesis: from poles to response}
\label{sec:modern}

The historical comparison suggests a common modern formulation that neither early tradition possessed in full. Start from a causal source-to-response operator rather than from a list of complex frequencies. For the simple stationary model of Eq.~\eqref{eq:master}, define in the physical half-plane
\begin{equation}
\vsub{G}{ret}(\omega)=\left[H-(\omega+\rmi0)^2\right]^{-1}
\end{equation}
in an appropriate weighted or distributional sense; the sign convention and the precise operator pencil must be adapted in frequency-dependent or coupled systems. For a physical source $|S\rangle$ and detector $\langle O|$, consider
\begin{equation}
\mathcal{A}(\omega)
=\langle O|\vsub{G}{ret}(\omega)|S\rangle.
\label{eq:response}
\end{equation}
Then analytically continue the response. Resonance poles, branch points, and regular background are parts of one analytic object.

This order resolves several persistent ambiguities. First, the pole position is separated from the pole strength. Second, the continuum is not treated as disposable background. Third, a non-self-adjoint eigenvalue is interpreted as a representation of the continued response rather than as the primitive definition of the physical system. Fourth, the same logic applies to both a quantum scattering experiment and a black-hole ringdown measurement.

In this formulation the older vocabularies become complementary rather than competing. The Siegert condition tells us which branch of the homogeneous equation defines a pole. The $S$ matrix tells us how asymptotic channels are connected on the real axis. The Feshbach operator tells us how eliminated channels dress a selected sector. Complex scaling represents the continued pole and continuum in an $L^2$ basis. The QNM language tells us how the same pole appears as a damped characteristic response of a black-hole exterior. The Green-function residue tells us how a specified source excites it.

The exchange should therefore go both ways. The goal is not to replace the QNM vocabulary by the Siegert vocabulary or vice versa. It is to know when a statement in one language has an exact translation in the other, when it has only an analogy, and when the geometry or channel structure makes the translation fail.

\section{Epilogue: a recent computational application}
\label{sec:epilogue}

The historical argument of this article does not depend on the author's recent calculations. As an epilogue, work by the present author and collaborators has applied finite-basis complex scaling to low-lying Schwarzschild and Reissner--Nordstr"om QNMs and has used a reference-subtracted continuum level density (CLD) for Schwarzschild--de~Sitter geometry \cite{Ogawa:2026veu,Ogawa:2026dSCSM}. Related work has also developed a response-level description of coalescing QNMs near exceptional points, using Riesz projectors and Laurent operators to characterize an isolated pole cluster without relying on the normalization or even the separate identification of individual modes \cite{Morikawa:2026RieszLaurent}. These studies are computational and spectral applications of the broader resonance framework discussed above, not claims to priority for the underlying black-hole resonance methods.

Their relevance to the present translation is operational. Together, they illustrate three distinct levels of the dictionary: isolated resonance poles, continuum spectral information, and residue-level source-to-observer response. A finite complex-scaled basis provides access to pole and continuum sectors with tools familiar from quantum resonance theory, while the Riesz--Laurent formulation becomes useful when individual modal residues cease to be well-conditioned near a pole coalescence. These quantities must nevertheless remain conceptually distinct. A reference-subtracted CLD is a global trace-level density, not by itself a retarded Green function, source-dependent excitation amplitude, or waveform; likewise, a representation of an isolated pole cluster should not be confused with the response measured for a specified source and observer. This epilogue therefore illustrates the practical use of the dictionary without making the recent calculations part of the historical argument.

\section{Conclusion}
\label{sec:conclusion}

Quantum resonance theory and black-hole QNM theory reached a common mathematical object along different historical paths. The quantum path began with radioactive decay and reaction resonances, moved through no-incoming boundary conditions and complex energies, and then confronted the problem that an outgoing decaying state is not an ordinary eigenvector of the original self-adjoint Hamiltonian. The black-hole path began with spacetime stability, reduced the perturbation equations to scattering barriers, discovered a characteristic damped response, and then developed a complex-frequency mode theory and Green-function decomposition.

The temporal ordering is clear. Kapur--Peierls posed an energy-dependent no-incident-wave problem in 1938, and Siegert supplied the self-consistent outgoing complex-energy pole state in 1939. Regge--Wheeler produced the black-hole barrier equation in 1957. Vishveshwara's 1968 dissertation then stated the horizon-ingoing/infinity-outgoing complex-frequency pole problem and connected it with resonance scattering; his 1970 stability paper carried the same construction into the journal literature, while his separate scattering paper displayed the characteristic ringing. Press gave that response its quasi-normal terminology in 1971. Chandrasekhar and Detweiler systematized the named frequency problem in 1975, and Leaver's 1986 response decomposition displayed pole, cut, and direct contributions in a form immediately recognizable to scattering theory.

The most striking date is 1971. Press was giving black-hole ringing its quasi-normal language at almost the same moment that Aguilar--Combes and Balslev--Combes were giving quantum resonances a rigorous complex-scaled spectral representation. The coincidence did not create a unified field. It reveals instead how two traditions, starting from different physical questions, converged on complementary descriptions of the same open-wave structure.

The later mathematical literature finally said the identification plainly. Black-hole QNMs are scattering resonances: poles of a continued response operator under the appropriate horizon and asymptotic conditions. The phrase should nevertheless be used with discipline. Historical genealogy is not mathematical equivalence, and a common pole does not erase the difference between a nuclear reaction and a curved spacetime.

The practical lesson is therefore a rule for translation. Begin with the causal or scattering problem, specify its asymptotic channels and time convention, identify the analytic quantity whose continuation defines the resonance, and only then choose a representation---Siegert state, QNM wave function, effective Hamiltonian, complex-scaled eigenvector, or resolvent pole. Once that order is respected, the two vocabularies cease to compete. They become two historically distinct languages for the same resonance.

\section*{Acknowledgments}

The perspective developed here grew out of parallel work on quantum resonance theory and black-hole perturbation theory. The author thanks collaborators and colleagues whose discussions made the mismatch of terminology between the two subjects increasingly visible.

The author would like to thank Izakaya YURURI for their hospitality.

The author used OpenAI's ChatGPT for literature discovery, structural
organization, and language drafting, and independently verified all cited
sources and historical claims.

\bibliographystyle{utphys}
\bibliography{ref}

\providecommand{\href}[2]{#2}\begingroup\raggedright\begin{thebibliography}{10}

\bibitem{KokkotasSchmidt:1999}
K.~D. Kokkotas and B.~G. Schmidt, ``Quasi-normal modes of stars and black holes,'' \href{http://dx.doi.org/10.12942/lrr-1999-2}{{\em Living Rev. Relativ.} {\bfseries 2} (1999) 2}.

\bibitem{DelaMadrid:2008GamowQNM}
R.~de~la Madrid, ``{The resonance amplitude associated with the Gamow states},'' \href{http://dx.doi.org/10.1016/j.nuclphysa.2008.08.003}{{\em Nucl. Phys. A} {\bfseries 812} (2008) 13--27}, \href{http://arxiv.org/abs/0810.0876}{{\ttfamily arXiv:0810.0876 [nucl-th]}}.

\bibitem{Ashida2020NonHermitian}
Y.~Ashida, Z.~Gong, and M.~Ueda, ``{Non-Hermitian physics},'' \href{http://dx.doi.org/10.1080/00018732.2021.1876991}{{\em Adv. Phys.} {\bfseries 69} no.~3, (2020) 249--435}, \href{http://arxiv.org/abs/2006.01837}{{\ttfamily arXiv:2006.01837 [cond-mat.mes-hall]}}.

\bibitem{Motohashi:2024fwt}
H.~Motohashi, ``{Resonant Excitation of Quasinormal Modes of Black Holes},'' \href{http://dx.doi.org/10.1103/PhysRevLett.134.141401}{{\em Phys. Rev. Lett.} {\bfseries 134} no.~14, (2025) 141401}, \href{http://arxiv.org/abs/2407.15191}{{\ttfamily arXiv:2407.15191 [gr-qc]}}.

\bibitem{Berti:2025hly}
E.~Berti, V.~Cardoso, G.~Carullo, {\em et~al.}, ``{Black hole spectroscopy: From theory to experiment},'' \href{http://dx.doi.org/10.1088/1361-6382/ae59e2}{{\em Class. Quant. Grav.} {\bfseries 43} (2026) 123001}, \href{http://arxiv.org/abs/2505.23895}{{\ttfamily arXiv:2505.23895 [gr-qc]}}.

\bibitem{KapurPeierls:1938}
P.~L. Kapur and R.~Peierls, ``The dispersion formula for nuclear reactions,'' \href{http://dx.doi.org/10.1098/rspa.1938.0093}{{\em Proc. Roy. Soc. Lond. A} {\bfseries 166} no.~925, (1938) 277--295}.

\bibitem{Siegert:1939}
A.~J.~F. Siegert, ``On the derivation of the dispersion formula for nuclear reactions,'' \href{http://dx.doi.org/10.1103/PhysRev.56.750}{{\em Phys. Rev.} {\bfseries 56} (Oct, 1939) 750--752}.

\bibitem{Regge:1957td}
T.~Regge and J.~A. Wheeler, ``{Stability of a Schwarzschild singularity},'' \href{http://dx.doi.org/10.1103/PhysRev.108.1063}{{\em Phys. Rev.} {\bfseries 108} (1957) 1063--1069}.

\bibitem{Vishveshwara:1968Thesis}
C.~V. Vishveshwara, \href{http://dx.doi.org/10.13016/M25X6S}{{\em The Stability of the {Schwarzschild} Metric}}.
\newblock PhD thesis, University of Maryland, College Park, 1968.
\newblock Chapter IV.B, pp. 51--58.

\bibitem{Vishveshwara:1970Stability}
C.~V. Vishveshwara, ``Stability of the {Schwarzschild} metric,'' \href{http://dx.doi.org/10.1103/PhysRevD.1.2870}{{\em Phys. Rev. D} {\bfseries 1} (1970) 2870--2879}.

\bibitem{Vishveshwara:1970Scattering}
C.~V. Vishveshwara, ``Scattering of gravitational radiation by a {Schwarzschild} black hole,'' \href{http://dx.doi.org/10.1038/227936a0}{{\em Nature} {\bfseries 227} (1970) 936--938}.

\bibitem{Press:1971QNM}
W.~H. Press, ``Long wave trains of gravitational waves from a vibrating black hole,'' \href{http://dx.doi.org/10.1086/180849}{{\em Astrophys. J. Lett.} {\bfseries 170} (1971) L105--L108}.

\bibitem{Aguilar:1971ve}
J.~Aguilar and J.~M. Combes, ``{A Class of Analytic Perturbations for One-Body {Schr\"odinger} Hamiltonians},'' \href{http://dx.doi.org/10.1007/BF01877510}{{\em Commun. Math. Phys.} {\bfseries 22} (1971) 269--279}.

\bibitem{Balslev:1971vb}
E.~Balslev and J.~M. Combes, ``{Spectral Properties of Many-Body {Schr\"odinger} Operators with Dilatation-Analytic Interactions},'' \href{http://dx.doi.org/10.1007/BF01877511}{{\em Commun. Math. Phys.} {\bfseries 22} (1971) 280--294}.

\bibitem{Newton:1960}
R.~G. Newton, ``Analytic properties of radial wave functions,'' \href{http://dx.doi.org/10.1063/1.1703665}{{\em Journal of Mathematical Physics} {\bfseries 1} no.~4, (07, 1960) 319--347}. \url{https://doi.org/10.1063/1.1703665}.

\bibitem{Simon:73}
B.~Simon, ``Resonances in n-body quantum systems with dilatation analytic potentials and the foundations of time-dependent perturbation theory,'' \href{http://dx.doi.org/10.2307/1970847}{{\em Annals of Mathematics} {\bfseries 97} no.~2, (1973) 247--274}. \url{http://www.jstor.org/stable/1970847}.

\bibitem{Moiseyev:1998gjp}
N.~Moiseyev, ``{Quantum theory of resonances: calculating energies, widths and cross-sections by complex scaling},'' \href{http://dx.doi.org/10.1016/S0370-1573(98)00002-7}{{\em Phys. Rept.} {\bfseries 302} no.~5-6, (1998) 212--293}.

\bibitem{Zworski:2017}
M.~Zworski, ``Mathematical study of scattering resonances,'' \href{http://dx.doi.org/10.1007/s13373-017-0099-4}{{\em Bull. Math. Sci.} {\bfseries 7} (2017) 1--85}, \href{http://arxiv.org/abs/1609.03550}{{\ttfamily arXiv:1609.03550 [math.AP]}}.

\bibitem{Sommerfeld:1912radiation}
A.~Sommerfeld, ``Die {Green}'sche funktion der schwingungsgleichung,'' {\em Jahresbericht der Deutschen Mathematiker-Vereinigung} {\bfseries 21} (1912) 309--353.

\bibitem{Schot:1992Sommerfeld}
S.~H. Schot, ``Eighty years of {Sommerfeld}'s radiation condition,'' \href{http://dx.doi.org/10.1016/0315-0860(92)90004-U}{{\em Historia Mathematica} {\bfseries 19} no.~4, (1992) 385--401}.

\bibitem{Gamow:1928ZPhys}
G.~Gamow, ``Zur quantentheorie des {Atomkernes},'' \href{http://dx.doi.org/10.1007/BF01343196}{{\em Z. Phys.} {\bfseries 51} (1928) 204--212}.

\bibitem{Gamow:1928Nature}
G.~Gamow, ``{The Quantum Theory of Nuclear Disintegration},'' \href{http://dx.doi.org/10.1038/122805b0}{{\em Nature} {\bfseries 122} (1928) 805--806}.

\bibitem{GurneyCondon:1928Nature}
R.~W. Gurney and E.~U. Condon, ``Wave mechanics and radioactive disintegration,'' \href{http://dx.doi.org/10.1038/122439a0}{{\em Nature} {\bfseries 122} (1928) 439}.

\bibitem{GurneyCondon:1929}
R.~W. Gurney and E.~U. Condon, ``Quantum mechanics and radioactive disintegration,'' \href{http://dx.doi.org/10.1103/PhysRev.33.127}{{\em Phys. Rev.} {\bfseries 33} (1929) 127--140}.

\bibitem{BreitWigner:1936}
G.~Breit and E.~P. Wigner, ``Capture of slow neutrons,'' \href{http://dx.doi.org/10.1103/PhysRev.49.519}{{\em Phys. Rev.} {\bfseries 49} (1936) 519--531}.

\bibitem{Hategan:2020}
C.~Hategan, R.~A. Ionescu, and H.~H. Wolter, ``{Siegert} state approach to quantum defect theory,'' \href{http://dx.doi.org/10.1140/epjd/e2020-100563-2}{{\em Eur. Phys. J. D} {\bfseries 74} (2020) 71}, \href{http://arxiv.org/abs/1607.07649}{{\ttfamily arXiv:1607.07649 [quant-ph]}}.

\bibitem{Friedrichs:1948}
K.~O. Friedrichs, ``On the perturbation of continuous spectra,'' \href{http://dx.doi.org/10.1002/cpa.3160010404}{{\em Commun. Pure Appl. Math.} {\bfseries 1} no.~4, (1948) 361--406}.

\bibitem{Eisenbud:1948}
L.~Eisenbud, {\em Formal Properties of Nuclear Collisions}.
\newblock PhD thesis, Princeton University, 1948.
\newblock Unpublished Ph.D. thesis.

\bibitem{Wigner:1955Delay}
E.~P. Wigner, ``Lower limit for the energy derivative of the scattering phase shift,'' \href{http://dx.doi.org/10.1103/PhysRev.98.145}{{\em Phys. Rev.} {\bfseries 98} (1955) 145--147}.

\bibitem{Smith:1960Lifetime}
F.~T. Smith, ``Lifetime matrix in collision theory,'' \href{http://dx.doi.org/10.1103/PhysRev.118.349}{{\em Phys. Rev.} {\bfseries 118} (1960) 349--356}.

\bibitem{Feshbach1958UnifiedI}
H.~Feshbach, ``{Unified Theory of Nuclear Reactions},'' \href{http://dx.doi.org/10.1016/0003-4916(58)90007-1}{{\em Annals of Physics} {\bfseries 5} no.~4, (Dec., 1958) 357--390}. \url{https://doi.org/10.1016/0003-4916(58)90007-1}.

\bibitem{Feshbach1962UnifiedII}
H.~Feshbach, ``{A Unified Theory of Nuclear Reactions. II},'' \href{http://dx.doi.org/10.1016/0003-4916(62)90221-X}{{\em Annals of Physics} {\bfseries 19} no.~2, (Aug., 1962) 287--313}. \url{https://doi.org/10.1016/0003-4916(62)90221-X}.

\bibitem{Peierls:1959Complex}
R.~E. Peierls, ``Complex eigenvalues in scattering theory,'' \href{http://dx.doi.org/10.1098/rspa.1959.0176}{{\em Proc. Roy. Soc. Lond. A} {\bfseries 253} no.~1272, (1959) 16--36}.

\bibitem{HumbletRosenfeld:1961}
J.~Humblet and L.~Rosenfeld, ``{Theory of Nuclear Reactions. I. Resonant States and Collision Matrix},'' \href{http://dx.doi.org/10.1016/0029-5582(61)90207-3}{{\em Nuclear Physics} {\bfseries 26} no.~4, (1961) 529--578}.

\bibitem{Kruskal:1960}
M.~D. Kruskal, ``Maximal extension of the {Schwarzschild} metric,'' \href{http://dx.doi.org/10.1103/PhysRev.119.1743}{{\em Phys. Rev.} {\bfseries 119} (1960) 1743--1745}.

\bibitem{Zel'dovich:1961}
Y.~B. Zel'dovich, ``{On the theory of unstable states},'' {\em Zh. \'Eksp. Teor. Fiz.} {\bfseries 39} (1960) 776. [English translation: Sov. Phys. JETP 12, 542 (1961)].

\bibitem{Berggren:1968zz}
T.~Berggren, ``{On the use of resonant states in eigenfunction expansions of scattering and reaction amplitudes},'' \href{http://dx.doi.org/10.1016/0375-9474(68)90593-9}{{\em Nucl. Phys. A} {\bfseries 109} (1968) 265--287}.

\bibitem{Zerilli:1970PRL}
F.~J. Zerilli, ``Effective potential for even-parity {Regge--Wheeler} gravitational perturbation equations,'' \href{http://dx.doi.org/10.1103/PhysRevLett.24.737}{{\em Phys. Rev. Lett.} {\bfseries 24} (1970) 737--738}.

\bibitem{Zerilli:1970wzz}
F.~J. Zerilli, ``{Gravitational Field of a Particle Falling in a {Schwarzschild} Geometry Analyzed in Tensor Harmonics},'' \href{http://dx.doi.org/10.1103/PhysRevD.2.2141}{{\em Phys. Rev. D} {\bfseries 2} (1970) 2141--2160}.

\bibitem{DadhichNayak:2024}
N.~Dadhich and R.~K. Nayak, ``{C. V. Vishveshwara} ({Vishu}) on the black hole trek,'' \href{http://dx.doi.org/10.1007/s12045-024-1735-4}{{\em Resonance} {\bfseries 29} (2024) 11--27}, \href{http://arxiv.org/abs/2402.11503}{{\ttfamily arXiv:2402.11503 [gr-qc]}}.

\bibitem{BessonJaramillo:2026}
J.~Besson and J.~L. Jaramillo, ``{Vishveshwara}'s waveform revisited: Insights from a {Keldysh} quasinormal mode expansion,'' \href{http://arxiv.org/abs/2608.11823}{{\ttfamily arXiv:2608.11823 [gr-qc]}}.

\bibitem{Simon1972BalslevCombes}
B.~Simon, ``Quadratic form techniques and the {Balslev-Combes} theorem,'' \href{http://dx.doi.org/10.1007/BF01649654}{{\em Commun. Math. Phys.} {\bfseries 27} no.~1, (1972) 1--9}.

\bibitem{Goebel:1972}
C.~J. Goebel, ``Comments on the ``vibrations'' of a black hole,'' \href{http://dx.doi.org/10.1086/180898}{{\em Astrophys. J. Lett.} {\bfseries 172} (1972) L95--L96}.

\bibitem{Teukolsky:1972my}
S.~A. Teukolsky, ``{Rotating {Black Holes}: Separable Wave Equations for Gravitational and Electromagnetic Perturbations},'' \href{http://dx.doi.org/10.1103/PhysRevLett.29.1114}{{\em Phys. Rev. Lett.} {\bfseries 29} (1972) 1114--1118}.

\bibitem{Teukolsky:1973}
S.~A. Teukolsky, ``{Perturbations of a Rotating Black Hole. I. Fundamental Equations for Gravitational, Electromagnetic, and Neutrino-Field Perturbations},'' \href{http://dx.doi.org/10.1086/152444}{{\em Astrophys. J.} {\bfseries 185} (1973) 635--647}.

\bibitem{ChandrasekharDetweiler:1975QNM}
S.~Chandrasekhar and S.~Detweiler, ``The quasi-normal modes of the {Schwarzschild} black hole,'' \href{http://dx.doi.org/10.1098/rspa.1975.0112}{{\em Proc. Roy. Soc. Lond. A} {\bfseries 344} no.~1639, (1975) 441--452}.

\bibitem{Leaver:1985ax}
E.~W. Leaver, ``{An Analytic representation for the quasi normal modes of Kerr black holes},'' \href{http://dx.doi.org/10.1098/rspa.1985.0119}{{\em Proc. Roy. Soc. Lond. A} {\bfseries 402} (1985) 285--298}.

\bibitem{Leaver:1986gd}
E.~W. Leaver, ``{Spectral decomposition of the perturbation response of the Schwarzschild geometry},'' \href{http://dx.doi.org/10.1103/PhysRevD.34.384}{{\em Phys. Rev. D} {\bfseries 34} (1986) 384--408}.

\bibitem{Price:1972I}
R.~H. Price, ``{Nonspherical Perturbations of Relativistic Gravitational Collapse. I. Scalar and Gravitational Perturbations},'' \href{http://dx.doi.org/10.1103/PhysRevD.5.2419}{{\em Phys. Rev. D} {\bfseries 5} (1972) 2419--2438}.

\bibitem{Ching:1995}
E.~S.~C. Ching, P.~T. Leung, W.~M. Suen, and K.~Young, ``Late-time tail of wave propagation on curved spacetimes,'' \href{http://dx.doi.org/10.1103/PhysRevLett.74.2414}{{\em Phys. Rev. Lett.} {\bfseries 74} (1995) 2414--2417}, \href{http://arxiv.org/abs/gr-qc/9410044}{{\ttfamily arXiv:gr-qc/9410044}}.

\bibitem{BachelotMotetBachelot:1993}
A.~Bachelot and A.~Motet-Bachelot, ``Les r\'esonances d'un trou noir de {Schwarzschild},'' {\em Ann. Inst. H. Poincar\'e Phys. Th\'eor.} {\bfseries 59} no.~1, (1993) 3--68. \url{https://www.numdam.org/item/AIHPA_1993__59_1_3_0/}.

\bibitem{SaBarretoZworski:1997}
A.~S\'a~Barreto and M.~Zworski, ``Distribution of resonances for spherical black holes,'' \href{http://dx.doi.org/10.4310/MRL.1997.v4.n1.a10}{{\em Math. Res. Lett.} {\bfseries 4} no.~1, (1997) 103--121}.

\bibitem{Vanzo:2004fy}
L.~Vanzo and S.~Zerbini, ``{Asymptotics of quasinormal modes for multihorizon black holes},'' \href{http://dx.doi.org/10.1103/PhysRevD.70.044030}{{\em Phys. Rev. D} {\bfseries 70} (2004) 044030}, \href{http://arxiv.org/abs/hep-th/0402103}{{\ttfamily arXiv:hep-th/0402103}}.

\bibitem{Dyatlov:2011QNM}
S.~Dyatlov, ``Quasi-normal modes and exponential energy decay for the {Kerr--de Sitter} black hole,'' \href{http://dx.doi.org/10.1007/s00220-011-1286-x}{{\em Commun. Math. Phys.} {\bfseries 306} (2011) 119--163}, \href{http://arxiv.org/abs/1003.6128}{{\ttfamily arXiv:1003.6128 [math.AP]}}.

\bibitem{Berti:2009kk}
E.~Berti, V.~Cardoso, and A.~O. Starinets, ``{Quasinormal modes of black holes and black branes},'' \href{http://dx.doi.org/10.1088/0264-9381/26/16/163001}{{\em Class. Quant. Grav.} {\bfseries 26} (2009) 163001}, \href{http://arxiv.org/abs/0905.2975}{{\ttfamily arXiv:0905.2975 [gr-qc]}}.

\bibitem{Cardoso:2009}
V.~Cardoso, A.~S. Miranda, E.~Berti, H.~Witek, and V.~T. Zanchin, ``Geodesic stability, {Lyapunov} exponents, and quasinormal modes,'' \href{http://dx.doi.org/10.1103/PhysRevD.79.064016}{{\em Phys. Rev. D} {\bfseries 79} (2009) 064016}, \href{http://arxiv.org/abs/0812.1806}{{\ttfamily arXiv:0812.1806 [hep-th]}}.

\bibitem{Ansorg:2016ztf}
M.~Ansorg and R.~Panosso~Macedo, ``{Spectral decomposition of black-hole perturbations on hyperboloidal slices},'' \href{http://dx.doi.org/10.1103/PhysRevD.93.124016}{{\em Phys. Rev. D} {\bfseries 93} no.~12, (2016) 124016}, \href{http://arxiv.org/abs/1604.02261}{{\ttfamily arXiv:1604.02261 [gr-qc]}}.

\bibitem{PanossoMacedo:2024nkw}
R.~Panosso~Macedo and A.~Zenginoglu, ``{Hyperboloidal approach to quasinormal modes},'' \href{http://dx.doi.org/10.3389/fphy.2024.1497601}{{\em Front. in Phys.} {\bfseries 12} (2024) 1497601}, \href{http://arxiv.org/abs/2409.11478}{{\ttfamily arXiv:2409.11478 [gr-qc]}}.

\bibitem{Nollert:1996rf}
H.-P. Nollert, ``{About the significance of quasinormal modes of black holes},'' \href{http://dx.doi.org/10.1103/PhysRevD.53.4397}{{\em Phys. Rev. D} {\bfseries 53} (1996) 4397--4402}, \href{http://arxiv.org/abs/gr-qc/9602032}{{\ttfamily arXiv:gr-qc/9602032}}.

\bibitem{NollertPrice:1999}
H.-P. Nollert and R.~H. Price, ``{Quantifying excitations of quasinormal mode systems},'' \href{http://dx.doi.org/10.1063/1.532698}{{\em J. Math. Phys.} {\bfseries 40} no.~2, (1999) 980--1010}, \href{http://arxiv.org/abs/gr-qc/9810074}{{\ttfamily arXiv:gr-qc/9810074}}.

\bibitem{Jaramillo:2020tuu}
J.~L. Jaramillo, R.~Panosso~Macedo, and L.~Al~Sheikh, ``{Pseudospectrum and Black Hole Quasinormal Mode Instability},'' \href{http://dx.doi.org/10.1103/PhysRevX.11.031003}{{\em Phys. Rev. X} {\bfseries 11} no.~3, (2021) 031003}, \href{http://arxiv.org/abs/2004.06434}{{\ttfamily arXiv:2004.06434 [gr-qc]}}.

\bibitem{Ogawa:2026veu}
S.~Ogawa, T.~Hirose, and O.~Morikawa, ``{Complex scaling approach to quasinormal modes of Schwarzschild and Reissner--Nordstr{\"o}m black holes},'' \href{http://dx.doi.org/10.1093/ptep/ptag160}{{\em Prog. Theor. Exp. Phys.} (2026) ptag160}, \href{http://arxiv.org/abs/2604.20442}{{\ttfamily arXiv:2604.20442 [hep-th]}}.

\bibitem{Ogawa:2026dSCSM}
S.~Ogawa, O.~Morikawa, and T.~Hirose, ``{Quasinormal modes and continuum response of de Sitter black holes via complex scaling method},'' \href{http://arxiv.org/abs/2605.03277}{{\ttfamily arXiv:2605.03277 [hep-th]}}.

\bibitem{Morikawa:2026RieszLaurent}
O.~Morikawa, S.~Ogawa, and T.~Hirose, ``{Riesz--Laurent representation of black-hole scattering and sourced response at exceptional points},'' \href{http://arxiv.org/abs/2608.14752}{{\ttfamily arXiv:2608.14752 [gr-qc]}}.

\end{thebibliography}\endgroup

\end{document}